\documentclass{emulateapj}
\slugcomment{For Submission to The Astrophysical Journal} 
\shorttitle{Mid-IR Spectra of SN 2003hv and SN 2005df}
\shortauthors{Gerardy et al.}
\newcommand{\nicrad}{\ensuremath{^{56}}Ni}
\newcommand{\cobrad}{\ensuremath{^{56}}Co}
\newcommand{\irnrad}{\ensuremath{^{56}}Fe}
\newcommand{\spitzer}{{\it Spitzer}}
\newcommand{\kms}{km~s$^{-1}$}

\begin{document}
\title{Signatures of Delayed Detonation, Asymmetry, and Electron Capture in 
the Mid-Infrared Spectra of Supernovae 2003hv and 2005df}
\author{Christopher L. Gerardy\altaffilmark{1}, W. P. S. Meikle\altaffilmark{1},
Rubina Kotak\altaffilmark{2}, Peter H\"oflich\altaffilmark{3}, 
Duncan Farrah\altaffilmark{4},  \\ Alexei V. Filippenko\altaffilmark{5},
Ryan J. Foley\altaffilmark{5}, Peter Lundqvist\altaffilmark{6}, 
Seppo Mattila\altaffilmark{7},
Monica Pozzo\altaffilmark{8},  \\ Jesper Sollerman\altaffilmark{9,10},
Schuyler D. Van Dyk\altaffilmark{11}
and J. Craig Wheeler\altaffilmark{12}
}
\altaffiltext{1}{Astrophysics Group, Blackett Laboratory, Imperial College 
London, Prince Consort Road, London SW7 2AZ, United Kingdom}
\altaffiltext{2}{European Southern Observatory, Karl-Schwarzschild-Str. 2, 
D-85748 Garching bei M\"unchen, Germany}
\altaffiltext{3}{Department of Physics, Florida State University, 315 Keen 
Building Tallahassee, FL 32306-4350}
\altaffiltext{4}{Department of Astronomy, Cornell University, Space
Sciences Building, Ithaca, NY 14853}
\altaffiltext{5}{Department of Astronomy, University of California,
Berkeley, CA 94720-3411}
\altaffiltext{6}{Stockholm Observatory, 133 36 Saltsj\"obaden, Sweden}
\altaffiltext{7}{Department of Physics and Astronomy, Queen's University
Belfast, Belfast, BT7 1NN, UK}
\altaffiltext{8}{Department of Earth Sciences, University College London, 
Gower Street, London, WC1E 6BT, United Kingdom}
\altaffiltext{9}{Dark Cosmology Center, Niels Bohr Institute, Copenhagen 
University Juliane Maries Vej 30, DK-2100 Copenhagen, Denmark}
\altaffiltext{10}{Stockholm Observatory, AlbaNova, SE-106 91 Stockholm, Sweden}
\altaffiltext{11}{\spitzer\ Science Center, California Institute of Technology,
Mail Code 220-6, Pasadena, CA 91125}
\altaffiltext{12}{Astronomy Department, University of Texas, Austin, TX 78712}
\begin{abstract}
We present mid-infrared (5.2--15.2 \micron) spectra of the Type~Ia supernovae
(SNe~Ia) 2003hv and 2005df observed with the {\it Spitzer Space Telescope}.
These are the first observed mid-infrared spectra of thermonuclear supernovae,
and show strong emission from fine-structure lines of Ni, Co, S, and Ar.  The
detection of Ni emission in SN 2005df 135 days after the explosion provides
direct observational evidence of high-density nuclear burning forming a
significant amount of stable Ni in a Type Ia supernova.  The observed emission
line profiles in the SN~2005df spectrum indicate a chemically stratified ejecta
structure.  The Ar and Ni lines in SN~2005df are both shifted to the red
relative to the Co emission, which appears to be nearly centered with respect
to the rest frame of the host galaxy.  The SN~2005df Ar lines also exhibit a
two-pronged emission profile implying that the Ar emission deviates
significantly from spherical symmetry.  The observed Ar lines can be reproduced
with either an almost circular ring-shaped geometry observed nearly edge-on, or
a prolate ellipsoidal emission geometry with a central hole observed nearly
pole-on.  The spectrum of SN~2003hv also shows signs of asymmetry, exhibiting
blueshifted [\ion{Co}{3}] which matches the blueshift of [\ion{Fe}{2}] lines in
nearly coeval NIR spectra.  Finally, local thermodynamic equilibrium abundance
estimates for the yield of radioactive \nicrad\ give $M_{^{56}Ni}\approx0.5
M_\sun$, for SN~2003hv, but only $M_{^{56}Ni}\approx0.13$--$0.22 M_\sun$ for
the apparently subluminous SN~2005df, supporting the notion that the luminosity
of SNe~Ia is primarily a function of the radioactive \nicrad\ yield.  

The chemically stratified ejecta structure observed in SN~2005df matches the 
predictions of delayed-detonation (DD) models, but is entirely incompatible with
current three-dimensional deflagration models.  Furthermore the degree that this layering 
persists to the innermost regions of the supernova is difficult to explain even
in a DD scenario, where the innermost ejecta are still the product of 
deflagration burning.  Thus, while these results are roughly consistent with a
delayed detonation, it is clear that a key piece of physics is still 
missing from our understanding of the earliest phases of SN~Ia explosions.

\end{abstract}

\section{Introduction}
There is general agreement about the basic nature of Type Ia supernovae 
(SNe~Ia): they are the thermonuclear explosion of degenerate white dwarfs 
(WDs), probably in some sort of close binary system undergoing some form of 
mass transfer, and exploding once a critical mass is reached, likely near the 
Chandrasekhar limit, at least for the majority of SN~Ia events. (For 
reviews, see \citealt{branch98,leibundgut01,nomoto03,hoflich03}.)  Still, many
of the details of this picture remain either unknown or are highly 
controversial. In particular, the initial conditions of the WD progenitor prior
to thermonuclear runaway, and the dynamical propagation of the nuclear burning
are both hotly debated topics of discussion.  

The recent growth in the availability of infrared (IR) observations shows much 
promise in addressing these issues. Infrared spectroscopy has several fundamental
advantages over UV/optical spectroscopy for work on SNe~Ia.  First, the strong 
spectral features in the IR are spread farther apart in wavelength,
avoiding most of the strong line-blending issues that complicate the analysis
of optical/UV observations.  Consequently, kinematic information can be extracted
directly from the observed line profiles without relying on a detailed spectral
synthesis calculation.  Second, IR lines tend to have a lower optical depth
than UV/optical lines which, combined with the reduced line blending, greatly
aids in the accurate measurement of abundances of iron-peak elements.  

This is particularly true of ground-state transitions in the mid-infrared
(MIR).  For these low-excitation lines the effects of stimulated emission are
very strong, dramatically reducing the optical depth.  Thus, these lines can
become optically thin quite rapidly, often while the electron densities are
still well above the critical density for collisional de-excitation.  As a
result, relatively accurate (and temperature insensitive) abundance estimates
can be made from MIR lines using just a local thermodynamic equilibrium (LTE)
approximation.  Finally, near-IR and mid-IR observations can be used to put
strong constraints on the formation of molecules and dust in the cooling
ejecta.

However, IR observations have one disadvantage. While the IR flux of SNe~Ia is
intrinsically fainter than in the optical, the foreground and background
emission is usually much larger.  In particular, foreground thermal emission
precludes MIR spectroscopy of all but the very nearest extragalactic SNe~Ia
from the ground, and to date, IR spectroscopic observations of SNe~Ia have been
limited to the near-infrared (NIR). With the launch of the {\it Spitzer Space
Telescope} \citep{werner04}, MIR spectroscopy is now possible, at least
for nearby ($D \lesssim 25$~Mpc) SNe~Ia.  In this paper we present the first
MIR spectra of SNe~Ia, obtained with the Infrared Spectrograph
\citep[IRS;][]{houck04} on the {\it Spitzer Space Telescope}: A
low signal-to-noise ratio (S/N) spectrum of SN~2003hv and the first of two planned
high S/N observations of SN~2005df.  Both were obtained as part of the GO
program of the Mid-Infrared Supernova Consortium (MISC).

In \S~2 we describe the observations and data reduction, and the results are 
shown in \S~3, including a detailed discussion of line identifications. In 
\S~4 we analyze the observed emission-line profiles using synthetic 
line profiles from simple models of the emission distribution. In \S~5, we 
use the observed nickel and cobalt line fluxes to infer abundances assuming
LTE level populations, and make estimates for the optical depth and electron
density.  Finally, we discuss and summarize our results in \S~6 and \S~7.

\section{Observations and Data Reduction}  
SN~2003hv was discovered at about 12.5 mag on 9.5 Sep 2003 (all dates UT) in
NGC 1201 by \citet{beutler03}, and classified as a SN~Ia on 10.3 Sep. 2003 by
\citet{dressler03}, closely resembling SN~1994D two days after maximum light.
The Lyon/Meudon Extragalactic Database
(LEDA)\footnote{\url{http://leda.univ-lyon1.fr}} gives a redshift corrected for
Virgo infall of 1494~\kms.  Assuming a Hubble constant $H_0=70$~\kms\
Mpc$^{-1}$, this implies a distance of 21.3 Mpc.  No light-curve data have yet
been published for SN~2003hv, but the reported discovery magnitude, Galactic
extinction of $A_V=0.05$ \citep{schlegel98}, and a 21.3 Mpc distance implies a
peak magnitude of $M_V \approx -19.2$.  Together with the reported initial
spectrum, this suggests that SN~2003hv was a relatively normal Type~Ia event.
Assuming that $B_{max}$ is indeed two days earlier than the initial spectrum
(8.5 Sep. 2003), and SN~2003hv is a normal SN~Ia, then a 19.5 day rise time
\citep{riess99} gives an estimated explosion date of 22 Aug. 2003.  (Note that
throughout this paper we will date the SN observation epochs with 
$t=0$ at the explosion, rather than the more common date of $B_{\rm max}$.)

SN~2005df was discovered visually in NGC 1559 on 4 Aug. 2005 by \citet{evans05},
and confirmed on 5 Aug. 2005 with a CCD image by \citet{gilmore05}.  It was
classified as a peculiar SN~Ia by \citet{salvo05}, who reported
that its optical spectrum on 5 Aug. 2005 resembled that of the moderately
subluminous Type~Ia SN~1986G a few days before maximum light.  As with
SN~2003hv, no light-curve data have yet been published, but observations
reported on
SNWeb\footnote{\url{http://www.astrosurf.com/snweb2/2005/05df/05dfMeas.htm}}
suggest that SN~2005df peaked near 12.3 mag around 18 Aug. 2005.

Several different distance estimates to NGC~1559 have been given. The LEDA
redshift of NGC~1559 corrected for Virgo infall is 1005~\kms, which yields a
distance of 14.4 Mpc for $H_0=70$~\kms\ Mpc$^{-1}$.  An expanding photosphere
method analysis of SN~1986L \citep{schmidt94,eastman96} gives a distance of
$16\pm2$~Mpc, and \citet{hamuy02} gives an estimate of $18.7\pm2.2$ Mpc based
on the \citet{tonry00} model of the peculiar velocity flow.  Adopting a peak
magnitude of 12.3, Galactic extinction $A_V=0.1$ mag \citep{schlegel98}, and no
host-galaxy extinction yields peak luminosities of $M_V\approx-18.5$ and
$-19.0$ mag for $D=14.4$ Mpc and 18.7 Mpc, respectively.  This suggests that
SN~2005df was subluminous by $\sim 0.5$--1 mag, relative to normal SNe~Ia
with $M_V\approx-19.5$ mag \citep{gibson00}. It is perhaps possible that some of
this is due to extinction in the host galaxy, but \textit{V}-band and
\textit{R}-band observations reported on SNWeb give $(V-R)\approx 0$ mag,
suggesting that SN~2005df is subject to relatively little reddening.  Assuming
$V_{max}$ occurred on 18 Aug., and adopting a 16 day rise time from 
delayed-detonation models appropriate to a moderately subluminous SN~Ia
\citep{hoflich02}, gives an estimated explosion date of 2 Aug. 2005.  

\spitzer\ observations of SN~2003hv consisted of a single set of short-low 
(5.2--15.2 \micron) IRS spectra, obtained as part of the GO1 program of the 
MISC.  Observations of SN~2005df were 
obtained via a low-impact target-of-opportunity (ToO) trigger as part of the 
MISC GO2 program.  The full set of ToO observations includes two epochs of 
imaging with the Infrared Array Camera \citep[IRAC;][]{fazio04}, two epochs of 
imaging with the blue peak-up array of the IRS (PUI), and 1 epoch of short-low  
spectroscopy with IRS.  In addition to these ToO observations, a second 
epoch of both short-low and long-low (14--38 \micron) IRS spectroscopy and a 
third set of IRAC/PUI imaging is scheduled for observation in GO3.  The full 
set of \spitzer\ data on SN~2005df will be analyzed in a later paper.  Here we 
will focus primarily on the first epoch of IRS spectroscopy.  

IRAC and PUI images of SN~2005df were obtained at three epochs, in Nov.
2005, Feb. 2006, and Aug. 2006, with the first two bracketing the IRS
spectral observations.  The imaging observations usually consisted of 30~s
exposures at 20 dither positions for each band, except for the Aug. 2006 PUI
images, for which four 30~s exposures were obtained for each of the 20 dither
positions.  

Aperture photometry was performed on the Post-Basic Calibrated Data (PBCD)
mosaic images using standard IRAF\footnote{IRAF is distributed by the National
Optical Astronomy Observatories, which are operated by the Association of
Universities for Research in Astronomy, Inc., under cooperative agreement with
the National Science Foundation.} routines and a circular aperture with a
three-pixel radius centered on the measured centroid of the detected point
source image.  The background was estimated using a circular annulus with inner
and outer radii of three and seven pixels, respectively.  The resulting fluxes
were multiplied by the aperture correction factors listed in version 3.0 of the
\textit{IRAC Data Handbook} and version 2.0 of the \textit{IRS Data Handbook}.
Photometric errors due to both photon statistics and background uncertainties
(estimated using the standard deviation in the background annulus) were
calculated, with the latter by far the dominant source of uncertainty.  For
images without a clear detection of the SN, an upper limit to the flux was
estimated using the uncertainty in the background emission at the location of
the SN.

IRS spectroscopy took place on 1 Sep. 2004 and 14 Dec. 2005 for SN~2003hv and 
SN~2005df, respectively.  Light curves for these events have not yet been 
published, so we will adopt our above estimates for the explosion dates 
placing the epochs of IRS observation at 375~d for SN~2003hv and 135~d for
SN~2005df.  These epochs are probably accurate to a few days, and this 
degree of timing uncertainty does not significantly affect our analysis.

IRS observations consisted of four sets of exposures: two nod positions spaced
19\arcsec\ apart in each of the two short-low slits (nominally 1st and 2nd
orders).  Fifteen 60~s exposures were obtained at each nod position in each
order, resulting in a total of 30~min integration time for each spectral order. 

Spectral reductions began with the Basic Calibrated Data (BCD) products from
the SSC pipeline (version S13.0.1). The fifteen exposures for each nod position
were combined using standard IRAF routines.  Background sky subtraction was
then achieved by subtracting the first nod position in each slit from the
second.  These subtracted frames were then cleaned using the contributed
IRSCLEAN\_MASK software
package\footnote{\url{http://ssc.spitzer.caltech.edu/archanaly/contributed/irsclean/}},
to remove hot pixels near the SN spectrum. One-dimensional (1-D) flux-calibrated spectra
were then extracted from the cleaned 2-D frames with the S13 version of
SPICE\footnote{\url{http://ssc.spitzer.caltech.edu/postbcd/spice.html}} using
the mask output from IRSCLEAN\_MASK and the uncertainty frames from the PBCD
data products.

The output of this process was four sets of flux-calibrated 1-D spectra, 
one from each nod position in each slit.  In addition, each of the extracted 
second-order spectra included data from the first-order ``bonus segment'' at the end 
of the detector.  So in total, the extracted short-low data consisted of 6 sets 
of partially overlapping 1-D spectra, two each covering 5.2--7.6~\micron\ 
(second order), 7.3--8.7~\micron\ (``bonus segment'') and 7.5--15.2~\micron\ 
(first order).  

These six 1-D spectra were then combined using custom software to make a
variance-weighted average of the data from each overlapping data set on a
bin-by-bin basis, using the error spectrum output by SPICE, and eliminating
any data points with abnormal flags.  In the overlap region, data from the
different spectral orders were combined by including flux from partially
overlapping bins in proportion to the size of the overlap compared with the size
of the bin (essentially assuming that the flux is spread evenly across a given
spectral bin).  This procedure was found to significantly increase the S/N of
the resulting combined spectrum compared with just a straight average of the
overlapping orders. This was particularly true in the region near 7 \micron,
where it is necessary to combine data near the ends of the individual spectral
orders because the S/N declines rapidly. 

\section{Results}
\subsection{SN 2005df Photometry}
\begin{figure}
\includegraphics[scale=0.30,angle=270]{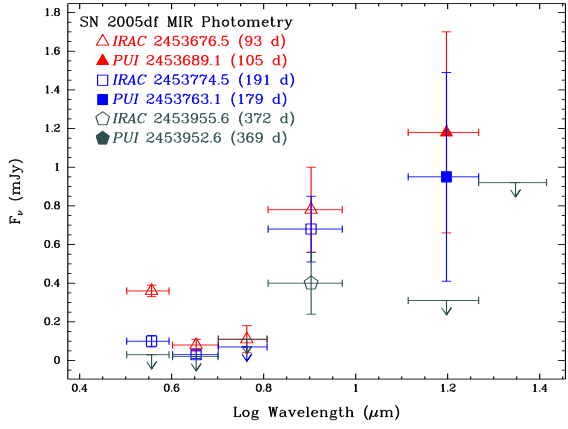}
\caption{Observed MIR broadband photometry of SN~2005df at three
epochs in Nov.\ 2005, Feb.\ 2006, and Aug.\ 2006.  The spectral energy
distribution shows little evidence of evolution except for the rapid fading
of the \textit{IRAC} Ch.\ 1 (3.6~\micron) flux between the first two epochs.
\label{fig1}}
\end{figure}

The broad-band photometric fluxes from the MIR images are presented in 
Table~\ref{mir_phot}\ and are plotted in Figure~\ref{fig1}.  The fluxes 
have relatively large uncertainties, particularly at the longer wavelengths, 
and these are, by far, dominated by the fluctuations in the background galaxy
emission near the SN.  In a subsequent paper we will try to address this
with a more careful extraction of the SN fluxes, but for now we will 
restrict ourselves to a fairly qualitative discussion of the photometry.

SN~2005df fades monotonically in all the observed bands, but for the most
part there is little evidence of significant evolution in the observed spectral
energy distribution (SED).  The one clear exception to this is the rapid
decline in the 3.6~\micron\ (IRAC Channel 1) flux, which is high in the
first epoch, but has fades by nearly a factor of 4 in the 100 days between the
first two epochs.  

The large 3.6~\micron/4.5~\micron\ ratio in the first epoch
suggests that strong line emission rather than continuum dominates the
3.6~\micron\ flux.  Potential candidate emission features in this band include
[\ion{Fe}{3}] 3.229~\micron, and [\ion{Co}{3}] 3.492~\micron, though
\textit{L}-band spectra of a SN~Ia around 100 days are needed to make a secure
identification.  Both of these lines have somewhat higher excitation
temperatures than the spectral features likely dominating the 8.0~\micron\ and
16~\micron\ bands, and thus the more rapid fading of the 3.6~\micron\ band
could be due to cooling in the ejecta.  A large number of [\ion{Fe}{2}] and 
[\ion{Co}{2}] lines also lie in the 3.6~\micron\ band but, as we will discuss 
in S~5.2, it seems likely that doubly ionized iron-peak species are dominant
in the ejecta at these epochs.

The flux is quite low at 4.5~\micron\ and 5.8~\micron\ at all epochs, the
latter consistent with the lack of emission seen at the blue end of the IRS
spectrum (Fig.~\ref{fig2}, \S~3.2).  This is in contrast to SNe~II,
which often show a strong flux excess at these wavelengths due to
emission from the fundamental ro-vibrational band of carbon monoxide
\citep[CO;][]{catchpole88,wooden89,kotak05,kotak06}.  The \textit{M}-band
\citep{catchpole88} flux from SN~1987A at 200~d (roughly the peak of CO
emission) would correspond to about 1.5--2 mJy at the distance (14.4--18.7 Mpc)
of SN~2005df.  All things being equal, the 4.5~\micron\ fluxes of SN~2005df
would thus imply an upper limit on the CO formation of $\lesssim 5$\% of
that formed in SN~1987A.  In practice, a more detailed analysis is needed to
account for radiation transport effects, and for the very different physical
conditions in the ejecta of SNe~Ia and SNe~II.

SN~2005df also shows significant emission at 8.0~\micron\ and 16~\micron.  Our
IRS spectrum suggests that the 8.0~\micron\ emission is dominated by strong
[\ion{Ar}{2}] and [\ion{Ar}{3}] emission, with some contribution from nickel
lines (see \S~3.2).  Possible lines contributing to the 16~\micron\
emission include [\ion{Co}{2}] 14.740, 15.647 and 16.30~\micron\, [\ion{Co}{3}]
16.391~\micron, [\ion{Co}{4}] 15.647~\micron\ (the ground state fine-structure
line of \ion{Co}{4}) and [\ion{Fe}{2}] 17.936~\micron.  Subsequent IRS
``Long-Low'' observations of SNe~Ia scheduled in GO3 will be used to make firm
identifications for the emission features in this band.

\subsection{SN 2005df Spectrum} 
\begin{figure*}
\includegraphics[angle=270,scale=0.65]{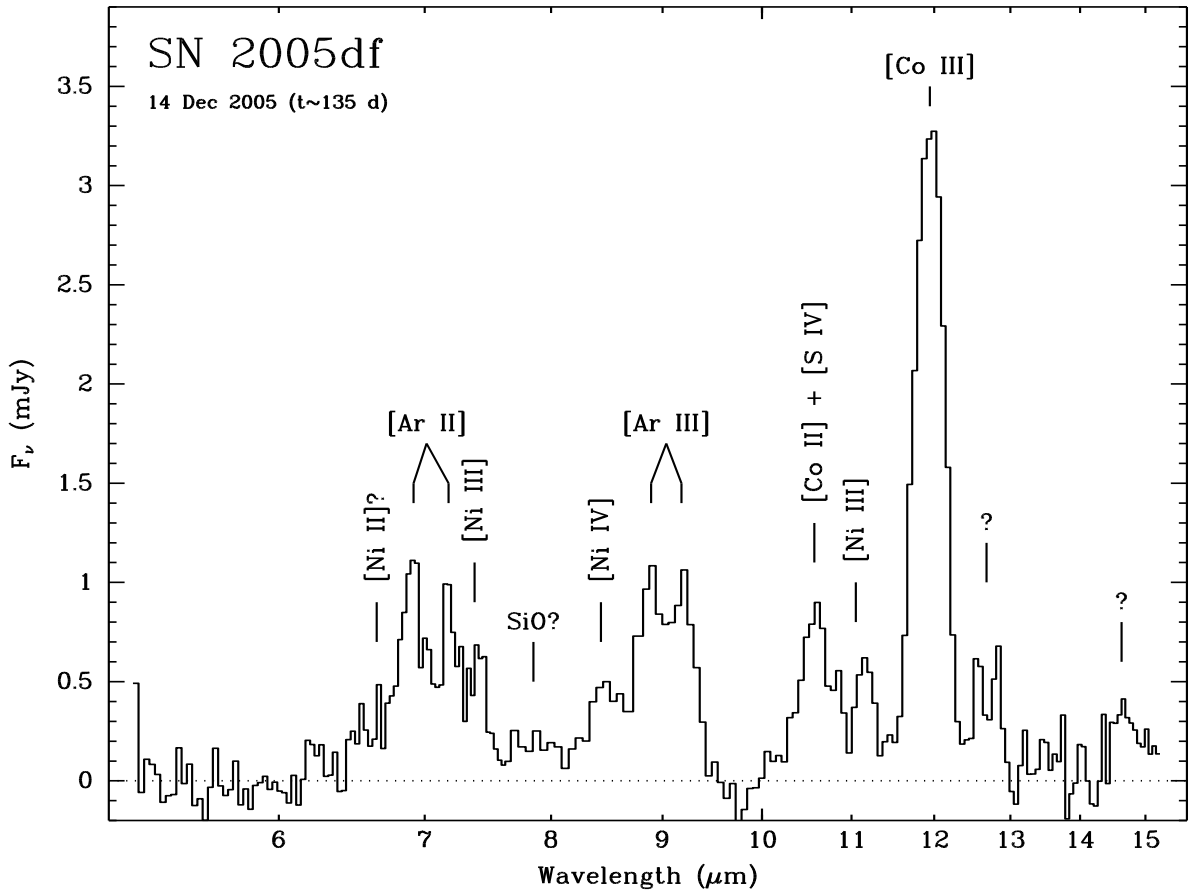}
\caption{The observed mid-infrared spectrum of SN~2005df.  Wavelengths are 
shown as vacuum coordinates in the observer's frame, and are plotted on a
logarithmic scale so that the observed line width is proportional to the
velocity line width throughout the large wavelength span.  See text for
discussion of line identifications.
\label{fig2}}
\end{figure*}

We begin our discussion of the MIR spectra with the second object observed,
SN~2005df, as the S/N of this spectrum is considerably higher than that of
SN~2003hv, and greatly aids the identification of spectral features.  The
reduced mid-infrared spectrum of SN 2005df on 14 Dec. 2005 is shown in
Fig.~\ref{fig2}.  The effective spectral resolution varies over the 
range 2500--4500~\kms, with the lowest resolution occurring at the blue end of each
order.

Line identifications were made by searching The Atomic Line
List\footnote{\url{http://www.pa.uky.edu/$\sim$peter/atomic/}} for strong
nebular transitions at or near the ground state. Candidate line transitions
were then further limited by the selection of atomic species with large
predicted abundances in models of SNe~Ia. Specifically, we
referred to the 1-D deflagration model W7 \citep{nomoto84}, and the 1-D
delayed-detonation (DD) model 5p02822.16 \citep[hereafter H02]{hoflich02} which
has a moderately subluminous synthetic light curve similar to that of SN~1986G.
Wherever possible, line identifications were tested by comparing the 
emission-line profiles of lines from the same atomic species.  The resulting line
identifications are marked in Figure~\ref{fig2}, and are listed in
Table~\ref{lines}.  

\subsubsection{Iron-Peak Elements}
\begin{figure}
\includegraphics[scale=0.39]{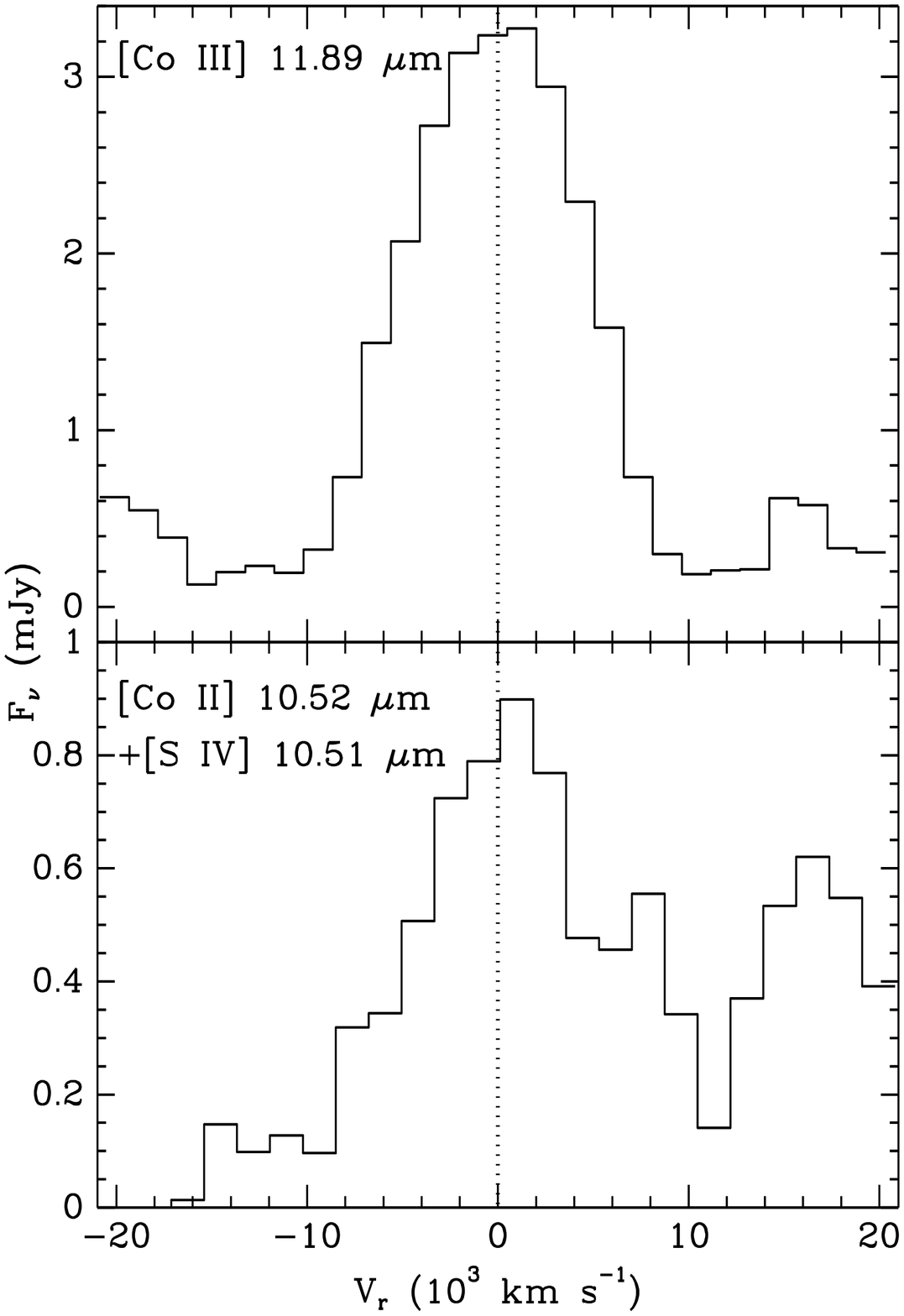}
\caption{The observed line emission from Co lines in the mid-IR spectrum
of SN~2005df on day $\sim 135$, plotted in velocity space relative to the 
rest frame of the host galaxy.
\label{fig3}}
\end{figure}

The spectrum of SN~2005df is, unsurprisingly, dominated by iron-peak elements,
showing strong Co emission and also several Ni lines.  Observed emission-line
profiles for the Co and Ni lines are shown in Figures~\ref{fig3} and \ref{fig4},
respectively.  Both of the Co lines appear to be fairly centered relative to
the host-galaxy rest frame.  The 11.89~\micron\ [\ion{Co}{3}] line is by far
the brightest feature in the spectrum and has a full-width at half-maximum (FWHM) of around 8000 \kms.
Although the S/N is much lower for the 10.52~\micron\ [\ion{Co}{2}] line, the
emission profile is not a great match to that of the [\ion{Co}{3}]
11.89~\micron\ line, as it appears to exhibit a broader base and, perhaps, a
slightly narrower core.  However, the [\ion{Co}{2}] 10.52 \micron\ emission
line is probably blended with [\ion{S}{4}].  In DD models (and W7), the S-rich
ejecta are predicted to have much larger radial velocities than the radioactive
Co-rich ejecta, making it tempting to suggest that the core of the 10.5
\micron\ feature is largely due to [\ion{Co}{2}] and the broader base due to
[\ion{S}{4}].  

In contrast to the Co emission, the Ni lines show a kinematic offset (of order
2000 \kms) relative to the host-galaxy rest frame. Unfortunately, two of the
three Ni lines are somewhat blended with the wings of the strong Ar emission
lines, making it difficult to compare the line profiles, though it appears the
[\ion{Ni}{4}] emission line is perhaps slightly broader than the [\ion{Ni}{3}]
lines.  The 11.00 \micron\ [\ion{Ni}{3}] line has a FWHM around 4000--5000
\kms.  Finally, we note that there is no convincing detection of the 6.63
\micron\ [\ion{Ni}{2}] line, although there is some possible excess emission at
the correct location just at the blue edge of the [\ion{Ar}{2}] emission
feature.  The rest frame wavelength of this [\ion{Ni}{2}] line is labelled in
Figure~\ref{fig2}. Note that the 1-pixel spike near the label is clearly just
noise.

\subsubsection{Intermediate-Mass Elements}
As well as the [\ion{S}{4}] emission likely contributing to the 10.52 \micron\
[\ion{Co}{2}] line, the spectrum contains further emission from moderate-mass
elements in the form of two Ar emission lines showing an unusual double-peaked
line profile (Fig.~\ref{fig5}).  As we will discuss in \S~\ref{arlines}, such a
profile strongly suggests a non-spherical but perhaps nearly axisymmetric
distribution of Ar emission.  Furthermore, as with the Ni emission, the Ar
lines appear to be redshifted relative to the rest frame of the host, with the
red and blue emission peaks appearing around $+6300$ and $-4300$ \kms,
respectively.  Given the dramatically different line profiles, it is difficult
to compare the width of the Ar lines with those of Co and Ni.  (This is further
complicated by the fact that both Ar lines are blended on at least one side
with other emission features.)  However, taking the 8.99 \micron\ line as the
cleaner of the pair, we can say that the full-width near zero-intensity (FWZI)
is at least 23000 \kms, compared to 16000 \kms\ for [\ion{Co}{3}] 11.89
\micron, and 8000 \kms\ for [\ion{Ni}{3}] 11.00~\micron.

\subsubsection{Molecular Emission?}
It is interesting to note that between the blended emission features near 7 and
9 \micron, the flux does not drop away to zero as it does shortward of 6
\micron, near 10 \micron, and longward of 13 \micron.  Instead the emission
levels out to about 0.2 mJy in the 7.7--8.1 \micron\ region.   This wavelength
region coincides with that of the fundamental ($\Delta v=1$) ro-vibrational
bands of silicon monoxide (SiO).  SiO molecular emission has been detected in
core-collapse supernovae (SN~1987A; \citealt{aitken88,wooden89,roche91};
SN~2005af; \citealt{kotak06}), but molecular emission has never been detected
in a thermonuclear supernova.   However, the identification of SiO in SN~2005df
should be considered speculative at best, as the excess emission could easily
be a blend of other unidentified faint features. 


\citet{hoflich95} studied the problem of molecule formation in detonation
models of thermonuclear supernovae and concluded that CO might form in
subluminous SNe~Ia, and SiO might form in SNe~Ia with a very low $^{56}$Ni
yield.  There is no evidence for CO fundamental emission at the blue end of the
spectrum, nor in the 4.5~\micron\ photometry of SN~2005df.  On the other hand,
the C/O-rich ejecta in DD models is much farther out than the layer containing
Si and O.  (In the 5p02822.16 model, the Si and O layers overlap at a radial
velocity of around 12500 \kms, while C is found predominantly beyond 19000
\kms.)  Since molecule formation rates are highly density dependent, it might
perhaps be possible to form SiO without forming CO. 

\subsubsection{Unidentified Features and Alternative Line IDs}
A few apparently real emission features in the spectrum 
of SN 2005df remain unidentified.  Two unresolved features with similar 
brightness appear to the red of the strong [\ion{Co}{3}] feature at 12.5 and 
12.8~\micron.  The redder of the two features might be residual background 
[\ion{Ne}{2}] 12.8 \micron\ emission. Indeed, examination of the raw 2-D spectra 
does show strong [\ion{Ne}{2}] emission, which might leave a small residual 
feature if the background is changing near the supernova.  However, we note 
that the 11.3 \micron\ [\ion{Ni}{1}] emission is much stronger in the 
background, and there is no comparable residual emission seen in the supernova 
spectrum.  It is perhaps possible that the residual 11.3 \micron\ 
[\ion{Ni}{1}] background could be masked by the [\ion{Ni}{3}] 11.00 \micron\ 
feature, though comparing the SN spectrum with the spectrum of nearby background
emission suggests that the two should be at least partially resolved.  In any 
case, [\ion{Ne}{2}] can only account for one of the two narrow features.  

Given the appearance of the Ar emission-line profile, it is perhaps tempting to
suggest that the two features are, in fact, a single feature with a forked
emission profile.  On the other hand, the spacing of the peaks does not match
those of the Ar lines, and indeed the entire feature is much narrower.  In sum,
no obvious ground-state line suggests itself for either the bluer line, or as
an identification for a single forked feature.  Similarly, an apparently broad emission
feature appears near the red edge of the spectrum at 14.6~\micron\ without an
obvious identification.

Finally, we note that it is possible to identify the emission features near 7.4
and 11.1 \micron\ as being due to [\ion{Ni}{1}] 7.507 and 11.308 \micron\
emission lines rather than [\ion{Ni}{3}].  These lines are two of the four
low-level fine-structure lines of [\ion{Ni}{1}], and a third, at 14.81 \micron,
could potentially match the 14.6 \micron\ feature. (The fourth at 12.00
\micron\ would be lost in the [\ion{Co}{3}] feature.)  However, the implied
$\sim 6000$ \kms\ blueshift of these three lines (see Fig.~\ref{fig6}), while
self consistent, is more difficult to understand in the context of current
SN~Ia explosion models. It is perhaps possible that such a high-velocity blob
of cold Ni could form in the context of a ``confined detonation'' scenario as
proposed by \citet{plewa04}.

\subsection{SN 2003hv Spectrum}
The reduced spectrum of SN 2003hv is shown in Figure~\ref{fig7} along with the
SN~2005df spectrum, allowing a direct comparison of the two.  The two spectra
appear to show many of the same features, although the S/N of the SN 2003hv
spectrum is much lower making definite identifications of the fainter features
difficult.  (Error bars are included in Fig.~\ref{fig8} in order to help
distinguish features and noise.)  The [\ion{Co}{3}] line which dominates the
SN~2005df spectrum is still visible in SN~2003hv, although it is much weaker
(both in absolute luminosity, and relative to other spectral features).
This weakening of the Co emission is not surprising, given that the SN~2003hv
spectrum was obtained at an age of more than twice that of the SN~2005df
spectrum, and indeed this supports the notion that most of the Co emission is
due to radioactive \cobrad.  

Like SN~2005df, SN~2003hv also exhibits strong [\ion{Ar}{2}] and [\ion{Ar}{3}]
emission features, and again they appear to exhibit complicated multi-peaked
emission-line profiles (Fig.~\ref{fig8}).  In SN~2003hv it appears that the
[\ion{Ar}{2}] and [\ion{Ar}{3}] line profiles may differ more significantly
than in SN~2005df, with [\ion{Ar}{2}] exhibiting narrow emission peaks at high
velocities not seen in [\ion{Ar}{3}].  The unidentified feature(s) near 12.5
\micron\ may also be present in the SN 2003hv spectrum, though there is no
convincing detection of the 12.8 or 14.6 \micron\ features.

It is possible that apparent excess emission near 11 \micron\ in the SN 2003hv
spectrum might be [\ion{Ni}{3}] or [\ion{Ni}{1}] as well, though the S/N of
this feature is very low.  However, both of these identifications are
upper-level transitions, and the ground-state transitions of neither
[\ion{Ni}{3}] nor [\ion{Ni}{1}] are seen near 7 \micron.  The red peak of the
[\ion{Ar}{2}] profile at 7.24~\micron\ (appearing near +9000 \kms\ in 
Fig.~\ref{fig8}) is not consistent with being either [\ion{Ni}{3}] 7.349
\micron\ at the same velocity as the 11 \micron\ emission identified as
[\ion{Ni}{3}] 11.002, nor [\ion{Ni}{1}] 7.507 \micron\ with the 11 \micron\
emission identified as [\ion{Ni}{1}] 11.308 \micron.  That there is no
detection of an appropriate 7 \micron\ line in SN 2003hv suggests that the 11
\micron\ feature may be just noise.  (On the other hand, the [\ion{Ni}{3}]
7.349 \micron\ line in SN~2005df is much weaker than expected compared to
[\ion{Ni}{3}] 11.002 \micron, perhaps because one or both features are
optically thick; see \S~\ref{nickel_abundance}.  It should also be noted that
the region near 7.4 \micron\ is where the edges of the three spectral orders
overlap, and the noise in this region may be underestimated.) Weak features
near 6.6 \micron\ and 8.5 \micron\ may be [\ion{Ni}{2}] 6.636 \micron, and
[\ion{Ni}{4}] 8.405 \micron, respectively, though again the S/N of these
putative features is quite low.

\section{Emission-Line Profiles\label{lineprofs}}
The line profiles of the resolved and relatively well-isolated emission features
in SN 2005df can be used to probe the chemical and ionization structure in the
ejecta.  To this end, we constructed model emission-line profiles based on 
relatively simple models of the kinematic distribution of emission.  
The models contain no explicit radiation transfer or radiative emission
calculations, but merely calculate the line profile for a given emissivity
distribution.  In effect, we are using these models to infer the geometry of the
emission distribution from the observed line profiles.  The details of the 
line-profile calculations are given in the Appendix.

\subsection{Nickel}
\begin{figure}
\includegraphics[scale=0.39]{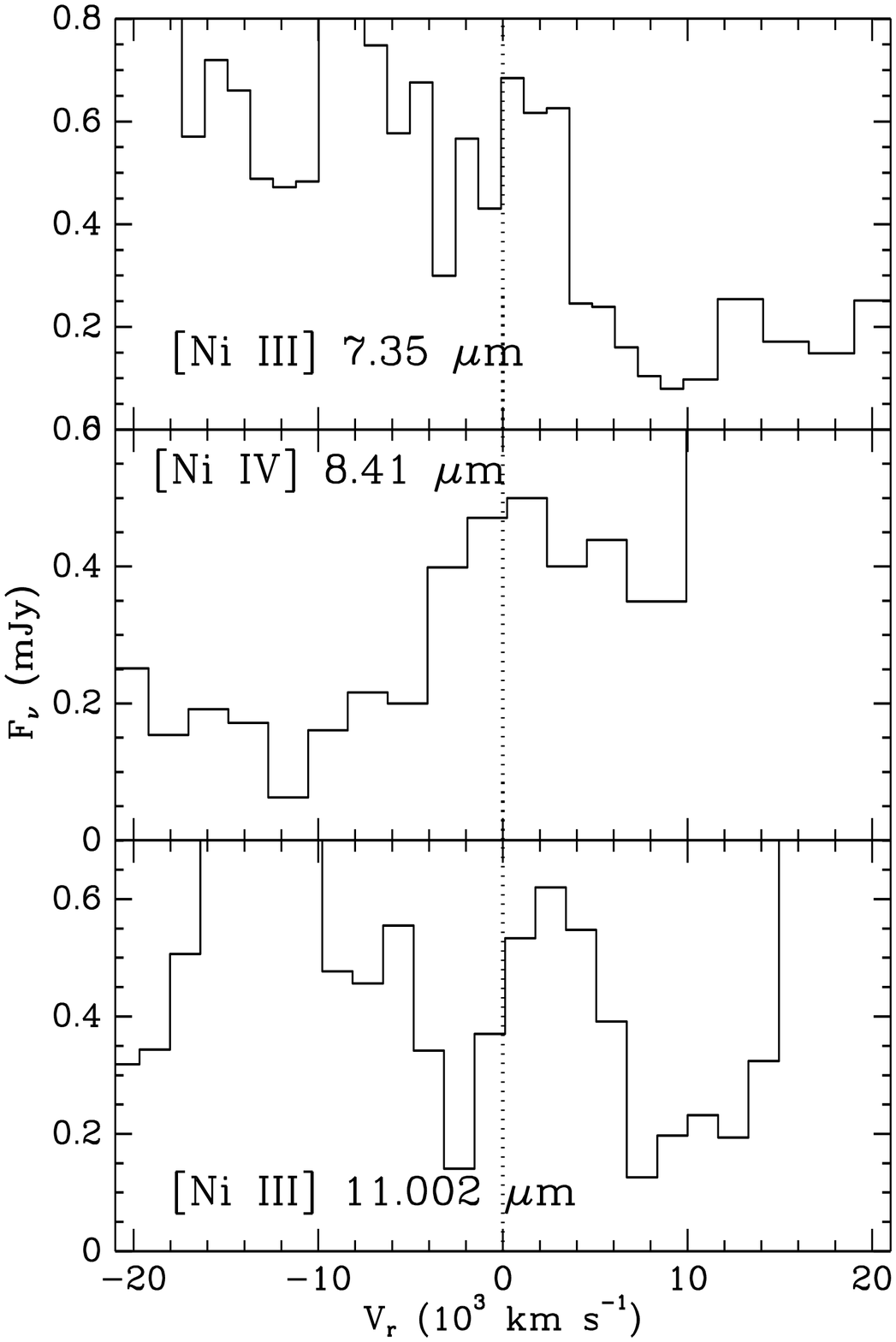}
\caption{The observed line emission from Ni lines in the mid-IR spectrum
of SN~2005df on day $\sim 135$, plotted in velocity space relative to the
rest frame of the host galaxy.
\label{fig4}}
\end{figure}

\begin{figure}
\includegraphics[scale=0.39]{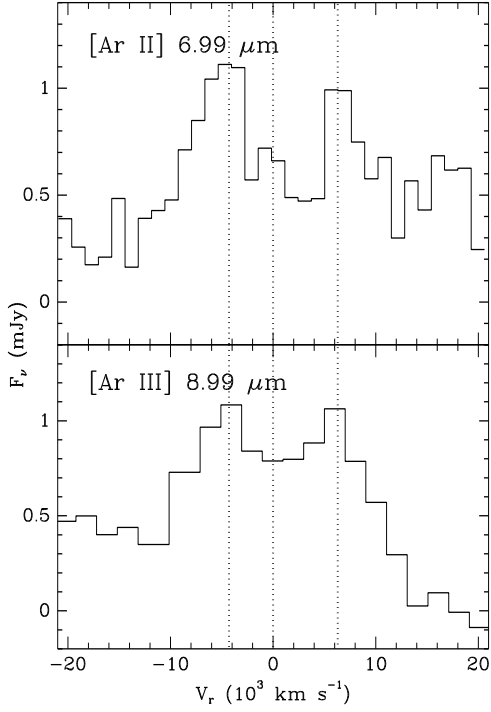}
\caption{The observed line emission from Ar lines in the mid-IR spectrum
of SN~2005df on day $\sim 135$, plotted in velocity space relative to the
rest frame of the host galaxy.  The central vertical dotted line indicates
emission at rest relative to the host, while the outer lines indicate the
approximate positions of the peaks in the Ar emission at $-4300$ and +6300
km~s$^{-1}$.
\label{fig5}}
\end{figure}

\begin{figure}
\includegraphics[scale=0.39]{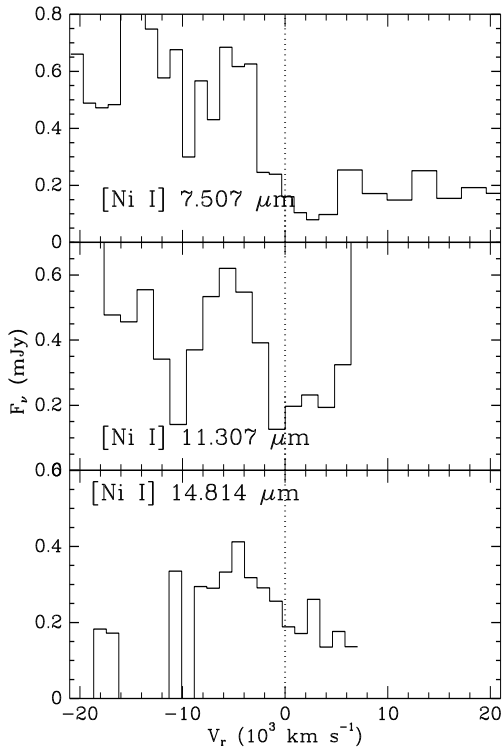}
\caption{Observed velocity profiles of three emission features in the 
mid-IR spectrum of SN~2005df on day $\sim 135$ that could 
potentially be consistent with [\ion{Ni}{1}] emission at high velocity.
\label{fig6}}
\end{figure}

\begin{figure*}
\includegraphics[angle=270,scale=0.65]{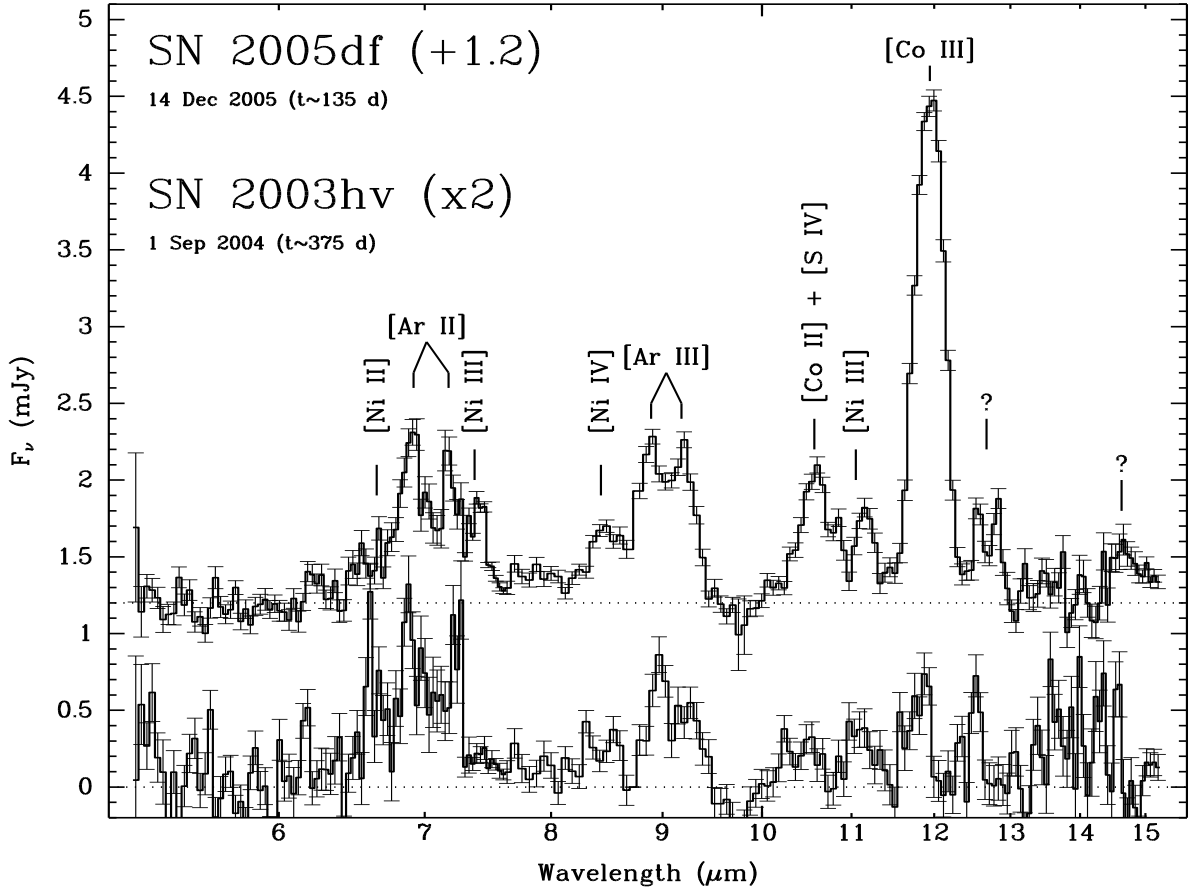}
\caption{The observed mid-IR spectrum of SN~2003hv compared with that of 
SN~2005df.  Wavelengths are 
shown as vacuum coordinates in the observer's frame, and are plotted on a
logarithmic scale so that the observed line width is proportional to the
velocity line width throughout the large wavelength span. Error bars
(based on the output error spectra from the individual order extractions, 
propagated through the weighted averaging of the orders)
are included to help distinguish noise from features in the low S/N spectrum of
SN~2003hv.  See text for discussion of line identifications.
\label{fig7}}
\end{figure*}

\begin{figure}
\includegraphics[scale=0.39]{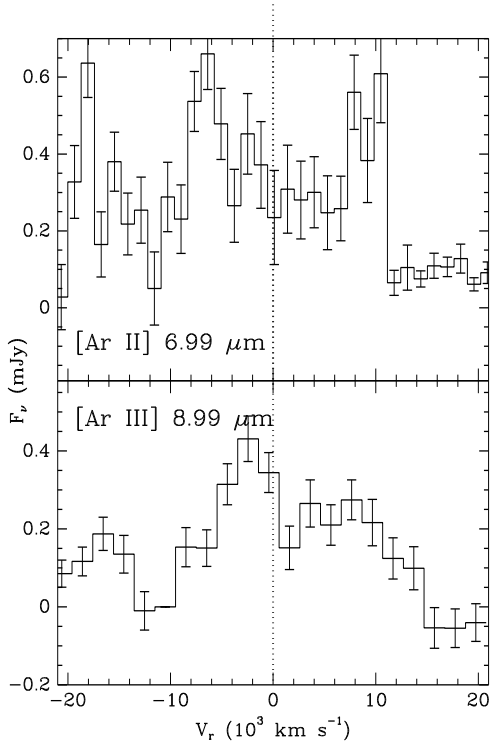}
\caption{The observed line emission from Ar lines in the mid-IR spectrum
SN~2003hv on day $\sim 375$, plotted in velocity space relative to the rest frame of the host
galaxy.  The central vertical dotted line indicates emission at rest relative
to the host. 
\label{fig8}}
\end{figure}

With a FWZI $\approx 8000$ \kms, the nickel lines in SN 2005df
(Fig.~\ref{fig4}) are marginally resolved.  Furthermore, two of the three
features are blended with the wings of the strong Ar emission lines.  At best
it can be said that the 11 \micron\ [\ion{Ni}{3}] line is consistent with the
parabolic line profile of a sphere of constant-density emission.   The
observed line centers of these features suggest that the Ni emission is
redshifted by roughly $\sim 2000$ \kms, relative to the rest frame of the host
galaxy.

\subsection{Cobalt}
\begin{figure}
\includegraphics[scale=0.30,angle=270]{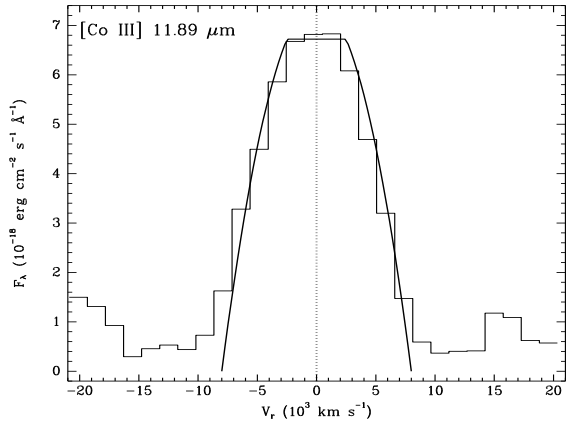}
\caption{The observed [\ion{Co}{3}] 11.89~\micron\ emission-line profile for 
SN~2005df on day $\sim 135$, compared to model line emission from a hollow uniform spherical
distribution of emission with an inner radius corresponding to a minimum 
velocity $v_{min}=2500$~\kms, and an outer radius corresponding to a maximum
velocity $v_{max}=8000$~\kms.
\label{fig9}}
\end{figure}

The 11.89~\micron\ [\ion{Co}{3}] line is also consistent with a simple
parabolic line profile, although the central pixels suggest a somewhat flatter
line core.  Indeed the line is a good match to the profile of uniform hollow
spherical emission distribution, with inner and outer radii corresponding to a
minimum velocity $v_{min}=2500$~\kms, and a maximum velocity
$v_{max}=8000$~\kms\ (see Fig.~\ref{fig9}).  Such a distribution is suggested by
the 5p02822.16 DD model, with the central hole corresponding to the
electron-capture zone.  The evidence for the central hole is relatively weak in
the mid-IR spectrum of SN~2005df due to the low spectral resolution, but it is
also observed in the line profile of the NIR [\ion{Co}{3}] 1.55~\micron\
feature around day 200 (see Fig.~\ref{fig10}; data from Gerardy et al., in
prep).  

That the flat line core is not seen in the [\ion{Fe}{2}] line at this time
probably indicates that gamma-rays from the radioactive \cobrad\ are still
being trapped in the core ejecta and that the energy deposition in the ejecta
is not yet completely local.  Little Co is produced in the electron-capture
zone, so the Co emission lines are flat-topped with both gamma-ray and positron
dominated energy deposition.  However, stable $^{54}$Fe is produced in the
electron-capture zone and will contribute to the Fe line profiles with
gamma-ray dominated non-local deposition.  There is no evidence for a
significant kinematic offset in the distribution of Co emission in SN 2005df.

In contrast, the [\ion{Co}{3}] 11.89~\micron\ line in SN 2003hv is
significantly blueshifted.  This blueshift is also seen in the [\ion{Fe}{2}]
lines of the 390~d NIR spectrum of SN~2003hv \citep{motohara06}.
Indeed, comparison of the two features in Figure~\ref{fig11} shows a generally
good match between the [\ion{Co}{3}] 11.89~\micron\ and [\ion{Fe}{2}]
1.64~\micron\ line profiles, although the S/N of the mid-IR spectrum is low
enough that the details of the [\ion{Co}{3}] are quite uncertain.  Still, the
correspondence between the two features lends strong support to the notion that
the late-time NIR [\ion{Fe}{2}] emission indeed traces the distribution of
radioactive ejecta as suggested by \citet{hoflich04}.

\subsection{Argon\label{arlines}}
The distinctive two-pronged emission-line profile of the Ar lines suggests
ring-shaped emission (see Fig.~\ref{example_profs}), but other emission
configurations are also possible.   This emission profile indicates a
significant lack of emission at low projected velocities.  However, simply
removing all the emission toward the center of spherically symmetric SN ejecta
is not sufficient to produce such a profile.  A hollow distribution will
produce a flat-topped profile, but will not create a central trough.  In fact,
in the limit where the Sobolev approximation holds (i.e., large velocity
gradients, and little continuum opacity), it is impossible to create the
observed argon profiles with any spherically symmetric homologously expanding
model.  Since spherically symmetric distributions can always be described as
the addition of concentric spherical shells, the emission-line profile from any
spherically symmetric distribution would always be a superposition of the
``top-hat'' emission-line profiles of spherical shells (see Appendix).

Thus, a two-pronged emission profile as seen in the Ar lines of SN~2005df
requires a significant deviation from spherical symmetry. These profiles can
potentially be formed by both a relatively circular, flattened geometry viewed
edge-on, or as a more prolate geometry viewed pole-on, with a large central
hole in both cases.  The velocities of the peaks are largely determined by the
central hole. The peak-to-trough contrast is mostly determined by the extent of
the deviation from spherical symmetry, with extreme geometries having deep
troughs, and more spherical geometries showing shallower troughs.  Indeed, the
observed line profile of [\ion{Ar}{2}] 6.99~\micron\ shows a much higher
peak-to-trough contrast than that of [\ion{Ar}{3}] 8.99~\micron, indicating
that the asymmetry is much stronger in the lower ionization species.

It is important to note that the observed argon line profiles only require an
asymmetric distribution of emission.  They do not necessarily imply a 
non-spherical distribution of argon in the ejecta, as the observed profiles 
could also be due to an asymmetric ionization structure or non-spherical 
energy deposition in the Ar-rich zone.  Indeed, as discussed above, the 
[\ion{Ar}{3}] emission does appear to be closer to spherical symmetry than
the [\ion{Ar}{2}] emission, and might suggest that the asymmetry is at least in
part an ionization effect.  However such and effect could also perhaps result 
from a non-spherical chemical or density distribution in the ejecta, and this
can not be ruled out with the analysis presented here.

Here we consider scenarios for both an edge-on ring-like symmetry and a
pole-on prolate symmetry, and consider a potential physical mechanism for
both cases.

\subsubsection{Ring Symmetry: the ``Magnetic Field Model''}
As the supernova ejecta become optically thin to gamma-rays, the energy 
deposition from the decay of \cobrad\ becomes dominated by positrons and fast
electrons.  The positrons are likely trapped quickly 
\citep[][although see references therein for opposing viewpoints]{sollerman04} but, as discussed by 
\citet[hereafter H04]{hoflich04}, the resulting fast electrons could travel 
quite a distance before thermalizing.  However, in order to explain the 
flat-topped [\ion{Fe}{2}] emission-line profiles observed in the late-time 
NIR spectrum of SN 2003du, the energy deposition from \cobrad\ needs to be 
trapped locally, leading H04 to suggest that a weak magnetic field was present
in the SN ejecta.  The efficiency of such magnetic trapping is not necessarily 
going to be spherically symmetric and might lead to an angular dependence in 
the energy deposition from positron decay, and thus in the ionization structure.

Such an effect would only be important at late epochs, when the ejecta become
optically thin to gamma-rays from the \cobrad\ zone.  In fact, the 135~d epoch
of the SN~2005df spectrum is probably a bit early for such an effect, and
indeed the 200 d NIR spectrum still shows strongly peaked [\ion{Fe}{2}]
emission (Gerardy et al. in prep; Fig.~\ref{fig11}).  However, the Ar appears at
much larger velocities and would become optically thin more rapidly than the
core iron.  Thus, it might be possible for positron deposition to be important
in the Ar zone while gamma-ray deposition still dominates in the Ni/Fe core.

As a mock-up of such a situation we considered emission models with a 
latitudinal angular profile given by 
\begin{equation}
\Psi(\psi)=\psi^\alpha e^{-\psi/\psi_0}
\end{equation}
and a power-law radial profile 
\begin{equation}
\rho(r)=r^{-\beta}
\end{equation}
truncated at a maximum radius $R_{max}=v_{max} t$.
The inner edge of the Ar emission was modelled by setting the flux to zero
within a spherical hole of radius $r_h$ and offset from the center along the
line of sight by $x_{off}$.  (In the formalism described in the Appendix, this
was achieved by setting the $\Phi(r,\psi,\phi)$ profile function to zero
anywhere within this sphere.)  The location of this hole is mainly constrained
by the peaks in the line profile, which appear to be asymmetric in both
velocity and strength.  The best fits to these peaks were found by using a hole
radius corresponding to a velocity of 5000 \kms.  Just offsetting the sphere by
1000 \kms\ to match the observed velocity asymmetry produces too much asymmetry
in the peak fluxes.  Instead, we found a better match by using a 500 \kms\
offset for the sphere, and then applying a further 500 \kms\ redshift to the
entire line profile, perhaps suggesting that the Ar emission as a whole is
somewhat redshifted, and that the central region is slightly more asymmetric.

We then tried to find a set of models that would match both Ar line profiles
with only a change in the angular distribution $\Psi(\psi)$. The parameter space
is large enough that such ``fits'' are by no means unique, but a reasonable
match (Fig.~\ref{fig12}) was found for $v_{max}=12000$ \kms, $\alpha=1.5$, 
$\beta=2$, and $\psi_0=0.25$\degr\ and $12$\degr\ for [\ion{Ar}{2}] and 
[\ion{Ar}{3}], respectively.  

The latitudinal profile functions for the two models are shown in
Figure~\ref{fig13}. The [\ion{Ar}{2}] model has a very flattened distribution
only covering a few degrees near the equator.  This is driven primarily by the
large peak-to-trough contrast, but there is some cross-talk with the other
parameters.  In particular, it is possible to fit the observed peak-to-trough
ratio with a wider angular dependence by using a narrower radial profile
(either by using a steeper radial power law or a lower maximum velocity),
although this tends to make it more difficult to fit the observed width of the
wings of the Ar profiles.  The narrowness of the angular profile apparently
required to reproduce the observed [\ion{Ar}{2}] line profile does cast some
doubt as to whether such an effect could be caused by the asymmetric magnetic
trapping described above.  This raises the question of whether other effects
(interaction with an accretion disk, for example) could create the required
extremely flattened equatorial ring-shaped emission.  Regardless, any proposed
physical mechanism must be able to simultaneously explain both the flattened
\ion{Ar}{2} and \ion{Ar}{3} distributions and the essentially spherical
\ion{Co}{3} distribution observed in SN~2005df.

We note that there is nothing special about the specific form used to describe
the emission here, and similar results can be obtained with other
flattened, relatively circular distributions.  For example, the line emission 
can also be modelled as oblate ellipsoids with a central spherical hole viewed 
edge-on.  This leads to similar results.  In particular, the observed 
peak-to-trough contrast in [\ion{Ar}{2}] again requires quite a flat ellipsoidal
distribution (major-to-minor axis ratios greater than 10:1). 

\subsubsection{Prolate Symmetry: the ``Off-Center DD Model''}
The argon emission-line profiles can also be reasonably reproduced by
a prolate emission distribution with a central hole, and viewed more pole-on
rather than edge-on as with the ring-shaped distribution.  The model line 
profiles shown in Figure~\ref{fig14}\ result from emission uniformly distributed
within prolate ellipsoids with a slightly off-center spherical hole in the 
middle.  

The emission regions are shown to scale in Figure~\ref{fig15}.
Both models have a major axis corresponding to an expansion velocity of 
12000 \kms\ in order to match the total width of the Ar features.
The minor axis was then chosen to match the observed peak-to-trough contrast.
The [\ion{Ar}{2}] model thus requires a narrower (more prolate) emission 
ellipsoid, corresponding to 6500 \kms, versus 8000 \kms\ for [\ion{Ar}{3}].

As with the ring profile, the central hole is largely determined by the 
velocity and relative strength of the two peaks and again has a radius 
corresponding to a velocity of 5000 \kms. But for these models the hole is 
shifted 1000 \kms\ to the back side of the SN envelope (rather than 500 \kms\ for
the ring models).  Again, as with the ring profiles, the entire emission line 
has been redshifted by a further 500 \kms.

Such an emission model is motivated by off-center DD models which produce an
asymmetric distribution in the innermost ejecta.  Figure~\ref{fig16}\ shows the
argon distribution from a 2-D hydrodynamic explosion model of an off-center
delayed detonation.  (This model is the same as that described by
\citet{fesen06}, and uses the same color scale as their Figure~7, with red being
the mass-fraction peak and blue being zero mass-fraction.)  

The DD model is generally more spherical than our ellipsoidal geometry, but
does show a pronounced crescent-shaped peak on one side.  This emission
geometry would only produce a single emission peak on a relatively broad and
flat distribution.  However, it is worth noting that the deflagration in this
model was calculated in 1-D, and therefore strictly maintained the spherical
symmetry of the core region prior to detonation.  It is perhaps possible that a
more realistic treatment of the deflagration might produce both a more
asymmetric chemical distribution, and could perhaps produce a second argon
peak, depending on the details of how the detonation propagates past the
deflagration core.   

\section{Abundances}
The Ni, Co, and Ar features observed in the SN 2005df spectrum all arise from 
low-level fine-structure transitions.  Most are ground-state transitions from
upper levels with excitation temperatures of 1000--2000 K, making them 
relatively insensitive to temperature.  (The one exception is the 
11.002~\micron\ [\ion{Ni}{3}] line, for which the lower level is the upper level
of the 7.349 \micron\ line, and the upper level has an excitation temperature 
near 3300~K. Consequently, this line is somewhat more sensitive to temperature.)
Furthermore, the low excitation temperature of the levels means that the 
stimulated emission correction is a large effect, which tends to keep the 
optical depth of these features relatively low.  Finally, in the early nebular
phase, the electron density in the ejecta can be well above the critical 
densities of the strong lines, which can make even simple LTE derived estimates
of the ejecta abundances relatively accurate.  Indeed, to some extent this can 
even be checked (see \S~5.5) by using the derived abundances and observed
line profiles to observationally infer the density in the ejecta.

In practice, a fully-NLTE nebular emission model is probably still needed for
robust abundance measurements, but such an analysis is beyond the scope of this
paper. We hope to include such an analysis in a subsequent paper using the
entire \spitzer\ dataset from SN~2005df as a constraint.  Here we will simply
use the observed line emission and an LTE approximation to estimate the
abundances.  Since we are extrapolating the LTE abundances from ground-state
transitions, NLTE corrections to these abundances will probably tend to
decrease the total mass, as the upper levels deplete relative to their LTE
populations.

Calculated abundances are given in Table~\ref{quantities}.  For each line 
included in the table we have inferred an observed mass from the number of atoms
in the upper level of the transition as implied by the line flux and the atomic
transition probabilities (values and references given in Table~\ref{lines}). 
Clearly this is an absolute lower limit to the abundances.  Assuming LTE level 
populations, the total mass of the ionization species can be calculated using 
the population of the upper level of the transition and the atomic
partition function. 

Note that in the following discussion we adopt the \citet{hamuy02} distance
of 18.7 Mpc to SN~2005df unless otherwise stated.  To scale the results to the 
16 Mpc distance of \citet{schmidt94} or the 14.4 Mpc distance of LEDA divide 
the masses and optical depths for SN~2005df by a factor of 1.4 or 1.7
respectively.

\subsection{Stable Nickel\label{nickel_abundance}}
\begin{figure}
\includegraphics[scale=0.39]{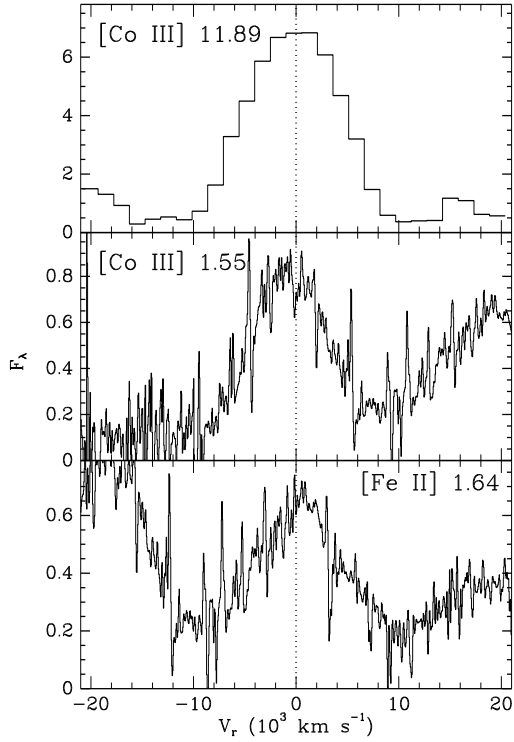}
\caption{The observed [\ion{Co}{3}] 11.89~\micron\ emission-line profile from
SN~2005df on day
135, compared with the near-IR emission-line profiles of [\ion{Co}{3}] 
1.55~\micron\ and [\ion{Fe}{2}] 1.64~\micron\ near day 200 (data from Gerardy
et al., in prep.) 
\label{fig10}}
\end{figure}

\begin{figure} 
\includegraphics[scale=0.39]{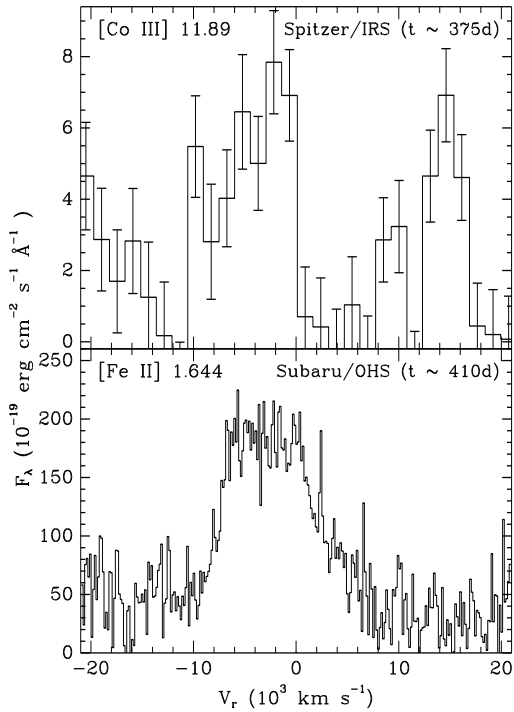} 
\caption{The observed
[\ion{Co}{3}] 11.89~\micron\ emission-line profile from SN~2003hv at 375~d,
compared with the 410~d near-IR emission-line profile of [\ion{Fe}{2}]
1.64~\micron\ (NIR data from \citealt{motohara06}).  
\label{fig11}}
\end{figure}

\begin{figure}
\includegraphics[scale=0.39,angle=0]{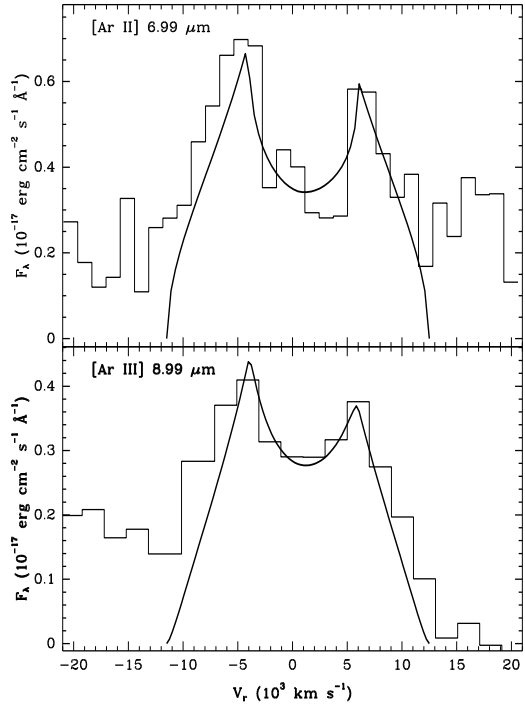}
\caption{
The observed Ar line profiles in SN 2005df on day $\sim 135$ compared with calculated emission-line profiles for an edge-on equatorial ring model.  The emission models for 
the two profiles differ only in the latitudinal profiles used (see 
Figure~\ref{fig13}), with the [\ion{Ar}{2}] emission confined to a narrow 
region near the equator and the [\ion{Ar}{3}] emission coming from a much larger
range of latitudes.  (See text for further details.)
\label{fig12}}
\end{figure}

\begin{figure}
\includegraphics[scale=0.30,angle=270]{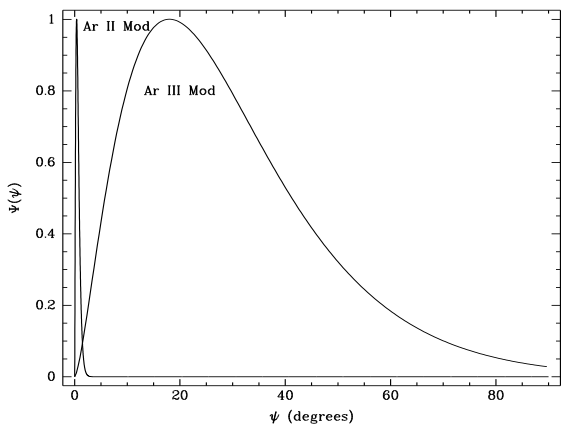}
\caption{
The latitudinal angular profile functions used for the models shown in 
Figure~\ref{fig12}.
\label{fig13}}
\end{figure}

\begin{figure}
\includegraphics[scale=0.39,angle=0]{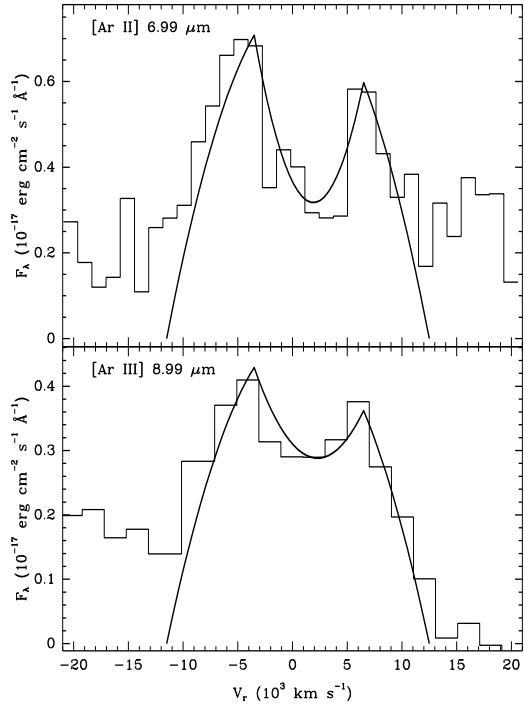}
\caption{
The observed Ar line profiles in SN 2005df on day $\sim 135$ compared with
calculated emission-line 
profiles for a pole-on prolate emission geometry with an off-center
spherical hole near the middle. (See text for details.)
\label{fig14}}
\end{figure}

\begin{figure}
\includegraphics[scale=0.3,angle=0,clip=true,trim=0bp 00bp 00bp 0bp]{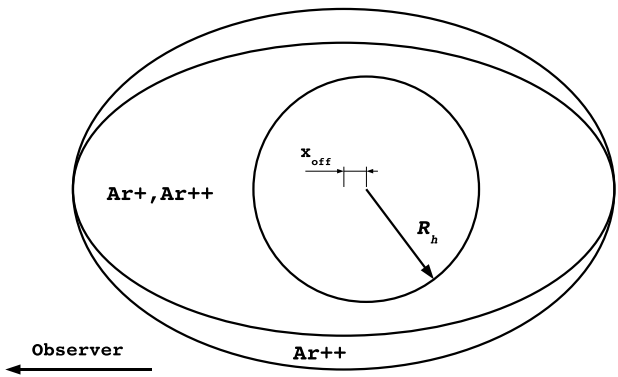}
\caption{Schematic of the model emission used to create the line profiles in
Figure~\ref{fig14}.   Emission is uniformly distributed in prolate ellipsoids
with a central spherical hole offset along the line of sight.  The figure is
to scale with the geometry used for the Ar line profiles in Figure~\ref{fig14}.
\label{fig15}}
\end{figure}

\begin{figure}
\includegraphics[scale=0.6]{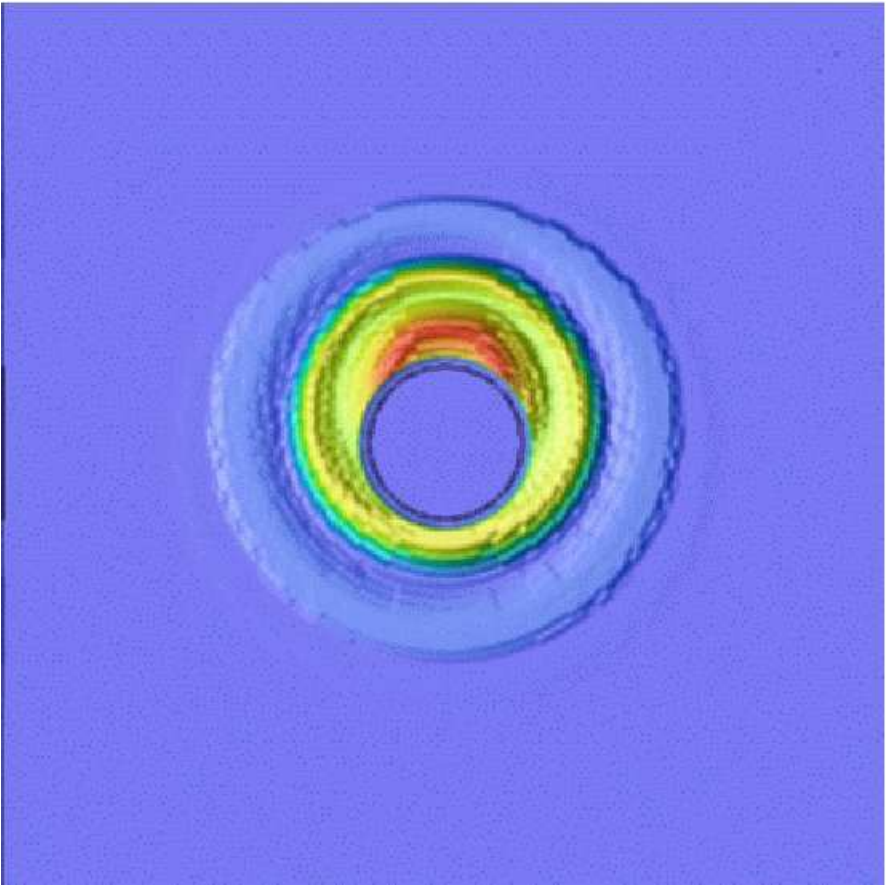}
\caption{Argon distribution from a 2-D model for an off-center delayed-detonation
model.  The model shown is the same as that described by \citet{fesen06}, and
the color scale is the same as their Figure 7, with red
corresponding to the mass-fraction peak, and blue corresponding to zero
mass-fraction.  The box dimensions correspond to velocities of $\pm 27500$~\kms.
The model shows a ``crescent-shaped" density peak which is perhaps qualitatively
similar to one lobe of our prolate Ar emission model (see text for discussion.)
\label{fig16}}
\end{figure}

For Ni, we use the partition functions of \citet{halenka01}, and calculate
masses for temperatures of 1000, 6000, and 10000 K.  Since the excitation
temperatures of these lines are low, the derived mass changes little for the higher
temperatures, but is dramatically increased at $T=1000$~K.  Such a temperature
is probably too low in the presence of local radioactive heating, but might
perhaps be realistic if the Ni is both spatially separated from the radioactive
Co and the Ni region is optically thin to gamma-rays from the radioactive zone.
The former is certainly suggested by the line profile analysis, but the the NIR
[\ion{Fe}{2}] line profile (Fig.~\ref{fig10}) suggests that the inner ejecta
are still trapping gamma-rays.  For the higher temperatures, the derived
abundances suggest a total Ni mass $M_{Ni}\approx (7-12) \times 10^{-3}
M_\sun$, somewhat low compared to the predictions of both W7 and the
5p02822.16 DD model, $M_{Ni}=6 \times 10^{-2} M_\sun$ and $7 \times 10^{-2} M_\sun$,
respectively.  However, the ratio of the observed [\ion{Ni}{3}] lines suggests
that they are optically thick, which may explain the discrepancy between the
derived and model Ni abundances.

Still, the implied Ni masses are much larger than can
be explained as radioactive \nicrad\ at such a late epoch (some 15+ half-lives
after the explosion).  Similarly, the derived Ni mass is at least an order of
magnitude larger than could be explained by a Chandrasekhar mass of primordial
nickel at solar abundance.  This strongly suggests that the observed Ni is an
electron capture product from high-density nuclear burning early in the 
explosion.  

\subsection{Radioactive Cobalt}
For the Co$^{2+}$ abundance, we use the partition function of \citet{halenka89},
for temperatures of 3000, 6000, and 10000~K.  In all three cases the derived
Co$^{2+}$ mass is quite similar, $M_{Co^{2+}} \approx 7 \times 10^{-2} M_\sun$.  

The number of \cobrad\ atoms at time $t$ is given by 
\begin{equation}
	N_2= \frac{N_0 \lambda_1}{\lambda_2 - \lambda_1} 
	\left(e^{-\lambda_1 t} - e^{- \lambda_2 t} \right),
	\label{decayeq}
\end{equation}
where $\lambda_1$ and $\lambda_2$ are the decay constants for \nicrad\ and
\cobrad, respectively, and $N_0$ is the total number of \nicrad\ atoms at $t=0$
\citep{colgate69}.  Assuming that Co$^{2+}$ is the dominant ionization species,
the 11.89~\micron\ [\ion{Co}{3}] line flux implies a relatively low \nicrad\ yield,
$M_{^{56}Ni} \approx 0.22 M_\sun$.  This is well below the $M_{^{56}Ni} =0.6
M_\sun$ prediction of W7, and other similar estimates of the \nicrad\ yield for
normal SNe~Ia.

For 6000 and 10000~K, the optical depth estimates for the 11.89~\micron\
line (see below) suggest that the line is relatively optically thin, and
correcting for the escape probability of 11.89~\micron\ photons only adds
another $\sim 20$\% to the inferred Co$^{2+}$ mass.  The implied optical depth
for 3000~K is 0.8, which is non-negligible.  Still, correcting for the escape
probability at $\tau=0.8$ only brings the \nicrad\ mass up to $\approx 0.32
M_\sun$.

It is perhaps possible that a significant mass of Co
is in a different ionization state.  However, the relative weakness of the
10.52~\micron\ [\ion{Co}{2}] line compared to the 11.89~\micron\ [\ion{Co}{3}]
line suggests that there is comparably little singly ionized Co.
Unfortunately, the ground state [\ion{Co}{4}] transition at 15.647 \micron\
lies just off the red edge of the spectrum.  \citet{rl95} estimate that the
Fe$^{3+}$/Fe ratio is less than 30\% based on UV observations of SNe~Ia
at an age of $\sim 1$ year and \citet{sollerman04} similarly find that the
contribution of Fe$^{3+}$ should be low at 300~d. Finally \citet{axelrod80}
finds Fe$^{3+}$/Fe $\approx$ 0.1--0.2 at 150 days. Hence, it seems unlikely that
the ionization could account for enough radioactive nickel to bring the
\nicrad\ yield up to match the $M_{^{56}Ni} \approx 0.6 M_\sun$ predicted by W7
and normal-luminosity DD models. 

We conclude that SN~2005df probably had a low yield of radioactive \nicrad\
by at least a factor of two compared to normal SNe~Ia.  This is even more
pronounced for the closer distance estimates for SN~2005df as both the \nicrad\
mass and the optical depth become smaller.  In contrast, the same calculations
for SN~2003hv (a spectroscopically normal SN~Ia) yield much larger \nicrad\
masses; $M_{^{56}Ni} \approx 0.47 M_\sun$ for $T=3000$~K and 6000~K, and
$M_{^{56}Ni} \approx 0.53 M_\sun$ for $T=10^4$~K.  Such masses are a good match
to the normal-luminosity DD models of H02 and are slightly less than the 0.6
$M_\sun$ of W7, though well within the errors from the line flux.  In fact, if
we assume $M_{^{56}Ni} =0.6 M_\sun$, these observations would imply that
Co$^{2+}$ is by far the dominant ionization state for Co in SN~2003hv.  

Thus, the \nicrad\ yield in SN~2005df appears low by a factor of $\sim2$ -- $4$
compared to both the model expectations for normal-bright SNe~Ia, and also to
the observed \nicrad\ yield from SN~2003hv, an apparently normal SN~Ia.
However, as we showed in \S~2, it also seems that SN~2005df was a
subluminous event, by roughly 0.5 mag ($D=18.7$ Mpc) to 1 mag ($D=14.4$ Mpc).
In fact, the derived \nicrad\ yields for these two distances are actually quite
close to the calculated yields for the subluminous DD models of HO2.  The
5p02822.14 model is about 1.5 mag underluminous, and has a predicted \nicrad\
yield of $M_{^{56}Ni} =0.15 M_\sun$.  This compares quite well to the results
assuming a 14.4 Mpc distance ($M_{^{56}Ni} =0.13 M_\sun$ and 1 mag
subluminous).  Similarly, the 5p02822.16 model is approximately 0.5 mag
subluminous and predicts a \nicrad\ yield of $M_{^{56}Ni}=0.27 M_\sun$ which
again compares relatively well to the observed results for an assumed 18.7 Mpc
distance ($M_{^{56}Ni} \approx 0.22 M_\sun$ and 0.5 mag subluminous).  Thus,
it seems that while the implied \nicrad\ yield is low compared with the nominal
expectations for normal-bright SNe~Ia, it is actually quite a decent match to
the predictions from subluminous DD models.  Furthermore, if these derived 
\nicrad\ abundances survive a more robust NLTE spectral analysis, and
if more detailed photometric analysis confirm that SN~2005df was indeed
subluminous, then these MIR observations provide strong direct support
for the notion that the luminosity of SNe~Ia is correlated with the \nicrad\
yield.  Note that \citet{stritzinger06}\ have also recently found strong 
evidence for this correlation based on the analysis of bolometric light-curves
and detailed modelling of the nebular spectra of nearby SNe Ia.

\subsection{Argon}
For the argon abundances, we calculated atomic partition functions for Ar$^+$
and Ar$^{2+}$ using the polynomial approximations of \citet{irwin81}.  Again,
we calculated LTE Ar$^+$ and Ar$^{2+}$ abundances for temperatures of 3000,
6000, and $10^4$~K, and the results are relatively insensitive to temperature,
yielding $M_{Ar^+}\approx 9\times 10^{-3} M_\sun$ and $M_{Ar^{2+}}\approx
1.5\times 10^{-2} M_\sun$.  The resulting Ar$^{2+}$/Ar$^+$ ratio is thus
approximately 2:1. However, the line-profile analysis suggests that the
[\ion{Ar}{3}] emission is coming from a larger region.  In fact, assuming the
emission is distributed as in our prolate emission model, the ionization is
approximately evenly split in the regions with Ar$^+$.  The total mass in
singly ionized and doubly ionized argon, $M_{Ar}\approx 2.4\times10^{-2} M_\sun$, is
close to the predictions of W7 and the 5p02822.16 DD model, which
predict $M_{Ar}\approx 2 \times10^{-2} M_\sun$ and $4\times10^{-2} M_\sun$, respectively.

\subsection{Optical Depth}
For homologously expanding ejecta, the Sobolev optical depth 
\citep[e.g.,][]{shu91} is given by 
\begin{equation}
\tau_\nu=\frac{n_l A_{u l}}{8 \pi} \frac{g_u}{g_l} \lambda^3 t \left(1 - 
e^{-h \nu/k T}\right),
\end{equation} where $n_l$ is the number density of atoms in the lower level and
$g_u$, $g_l$ are the statistical weights of the upper and lower levels, respectively.
Assuming an LTE relation between the upper and lower level populations given
by 
\begin{equation}
n_l=n_u \frac{g_l}{g_u} e^{(h \nu/k T)},
\end{equation}
the Sobolev optical depth becomes
\begin{equation}
\tau_\nu=\frac{n_u A_{u l}}{8 \pi} \lambda^3 t
\left(e^{h \nu/k T} - 1\right).
\end{equation}
From this, an LTE optical depth can be calculated (for a given temperature) 
from the total number of 
upper-level atoms implied by the line flux, converted to a number density by
a judicious assumption about the distribution based on the line-profile
analysis. 

The logic of such a calculation can be slightly circular, in that the total
number of emitting atoms is inferred from the observed flux assuming that the
line is optically thin.  This is then converted to a density to infer an
optical depth.  If the calculated optical depth is indeed low, then the
estimate from such an analysis is probably valid.  However, if the calculated
optical depth approaches unity (or beyond), then the actual number of emitting
atoms will be underestimated, and the actual optical depth will be higher,
perhaps significantly so.

For each of the abundance estimates in Table~\ref{quantities}, we have also
calculated an LTE optical depth.  For the Ni lines, we assumed that the Ni was
uniformly distributed in a sphere with a radius corresponding to an expansion
velocity of 4000 \kms.  For the Co lines, we also assumed a uniform
distribution, but this time in a hollow spherical volume with an inner radius
corresponding to 2500 \kms\ and an outer radius of 8000 \kms.  

This analysis implies that the Ni lines are optically thin at 6000 and
10000~K, but optically thick at 1000~K. The observed flux ratio of the two
[\ion{Ni}{3}] lines suggests that these lines may indeed be optically thick.
In principle, the [\ion{Ni}{3}] 11.002/7.349 \micron\ line ratio can be used as
a temperature probe (at least for low temperatures).  The observed line ratio,
$0.6\pm0.2$, is much larger than it should be.  \citet{bautista01} has
calculated the 11.00/7.35 \micron\ line ratio for a large range of densities
and temperatures, and the ratio is always less than 0.2.  While the observed
line ratio is not well constrained, making it consistent with the
\citet{bautista01} calculations would require a rather large change in the
observed line fluxes.  Indeed, taken at face value, the upper level populations
inferred from the line ratio imply inverted level populations for the 11.002
\micron\ line.  It seems more likely that the Ni lines are simply
optically thick.

The optical depth of the Co lines in SN~2005df is relatively low for all the 
abundance estimates, except for $T=3000$~K where it is beginning to 
become an important effect.  Still, even at 3000~K, the Sobolev escape 
probability for the [\ion{Co}{3}] 11.89~\micron\ photons is around 70\%,
so it is unlikely that a large amount of [\ion{Co}{3}] is being missed.  
If the temperature in the Co zone was as low as 1000~K, then the line would
be optically thick, but it seems unlikely that the temperature has become so 
low in the \cobrad\ region 135 days after the explosion.  
The optical depth of the [\ion{Co}{3}] 11.89~\micron\ line in SN~2003hv is 
negligible for all temperatures.

Optical depth calculations are slightly more complicated for the Ar features,
as the densities implied by the line emission depend on the assumptions made
about the emission distribution.  For the optical depths quoted in
Table~\ref{quantities}, we assumed the emission was distributed as in our
prolate line-profile model.  The resulting optical depths are small, suggesting
that the Ar lines are indeed optically thin.

\subsection{Electron Density}
One check on the robustness of the derived abundances would be to see if the 
implied electron density is above the critical density for the observed ions.
If it is, then the collisional processes will dominate and tend to drive the 
level populations toward LTE, and thus our LTE mass estimates would be 
relatively valid.

Unfortunately, there are almost no published collision strengths for 
doubly ionized cobalt, so we used the approximation
\begin{equation}
\Omega_{ij}=\omega g_i g_j,
\end{equation}
with $\omega=0.2$ \citep{axelrod80,graham87,bowers97}.  This yields a 
typical critical density $n_{crit}\approx 10^{6}$ cm$^{-3}$ for the strong
optical, NIR, and MIR [\ion{Co}{3}] transitions.   Published collision
strengths for Ar$^+$ and Ar$^{2+}$ \citep{pelan95,galavis95} give critical 
densities $n_{crit}\approx 3\times10^{5}$ cm$^{-3}$ for both of the Ar
transitions seen in SN~2005df and SN~2003hv.

For SN 2005df, the Co$^{2+}$ number density as calculated for the optical depth
estimates is $n_{Co^{2+}}\approx1.4\times10^{5}$~cm$^{-3}$.  Assuming that
Co$^{2+}$ is the dominant ionization species, and an epoch  
of 135 days, Equation~\ref{decayeq} implies that the total density of
\nicrad, \cobrad, and \irnrad\ ions in the Co-emitting zone is 
$n_{\nicrad}+n_{\cobrad}+n_{\irnrad}\approx4.5\times10^{5}$~cm$^{-3}$.  
In the 5p02822.16 DD model, the mass fraction of \nicrad\ in the \nicrad-rich
zone is $\sim 0.6$--0.8, so the total ion density in this region should 
be about 30\% higher than the density of radioactive ejecta products, 
$n_{ion} \approx 5.5\times10^5$~cm$^{-3}$.  

The total electron density will depend on the average ionization state of ions
in the radioactive zone.  Ionization models for SNe~Ia suggest that Fe$^+$
should dominate in the \cobrad\ zone at an age of around 300 days
\citep{rl95,liu97,sollerman04}.  This would give an electron density $n_e
\approx n_{ion}$, and thus even this relatively low ionization is enough to put
the electron density above the critical densities for the Ar lines, and near
the critical densities for Co.  Furthermore, at this earlier epoch the
ionization is likely to be higher, which could drive the electron density over
the critical density.  

This suggests that the LTE approximation is not too bad, at least for the Co
and Ar abundance in SN~2005df, and that NLTE level populations are unlikely to
explain to low \nicrad\ abundance in the supernova.  The same
calculation for SN~2003hv yields a significantly lower ion density
$n_{ion}\approx5\times10^{4}$~cm$^{-3}$, and suggests that the NLTE corrections
to the abundance may not be small in the older supernova.

\section{Discussion}
One of the most controversial topics of discussion in SN~Ia physics is the 
question of the exact nature of the thermonuclear burning front.  In particular,
does it propagate primarily as a subsonic deflagration front or as a near-sonic
or supersonic detonation? The distinction is important because WDs undergoing 
deflagration have time to react to the release of nuclear energy and evolve 
dynamically during the supernova explosion, while detonations proceed through
the star too rapidly for dynamical evolution to occur. 

Pure detonation models have long been ruled out as they burn the WD progenitor
almost entirely to iron-peak elements \citep[e.g.,][]{nomoto82} and cannot 
reproduce the observed intermediate-mass spectral features of Si, S, Mg and O 
typically seen in SN Ia spectra \citep[][and references therein]{filippenko97}. 
In contrast, 1-D pure deflagration models such as W7 result in a spectrum of 
burning products ranging from iron-peak, to intermediate-mass
elements, and also leave a significant portion of the C/O progenitor unburned
\citep[e.g.,][]{nomoto84}.  

A similar spectrum of burning products results from the delayed-detonation
scenario \citep{khokhlov91,yamaoka92}, where the burning begins as a
deflagration but quickly transitions to a detonation.  The deflagration serves
to pre-expand the WD, lowering the density of the outer regions of the star so
that the resulting detonation produces a spectrum of burning products and not
simply Fe-peak elements.  Since the detonation propagates too fast for the
formation of convective turbulence, the WD undergoes essentially no
hydrodynamic evolution during that stage of the burning, and thus a radial
chemical structure is produced even in multi-dimensional models
\citep{gamezo05}. 

While 1-D deflagration models produce a radially stratified structure, 
3-D deflagration models have shown that the burning is subject to instabilities 
which result in the formation of large-scale convective plumes and macroscopic 
mixing of the ejecta \citep[e.g.,][]{gamezo04,roepke06}.  As a result, 3-D 
deflagration models predict a much more chemically homogeneous ejecta structure.

In the two SNe observed here, we find strong evidence for a layered chemical
structure like that seen in DD models (or 1-D deflagrations like W7).  The line
profiles in SN~2005df suggest a clear radial progression from Ni ($v \lesssim
4000$~\kms), to Co ($2500 \lesssim v \lesssim 8000$~\kms) to Ar ($5000 \lesssim
v \lesssim 12000$~\kms).  The detection of Ni in SN~2003hv is less clear, but
certainly the Ar lines are again much broader than the observed Co line in this
event.  Thus the MIR spectra come down quite clearly against the sort of
large-scale mixing predicted by the 3-D deflagration models.

\begin{figure*}
\includegraphics[scale=0.95,clip=True,trim=0bp 440bp 0bp 00bp]{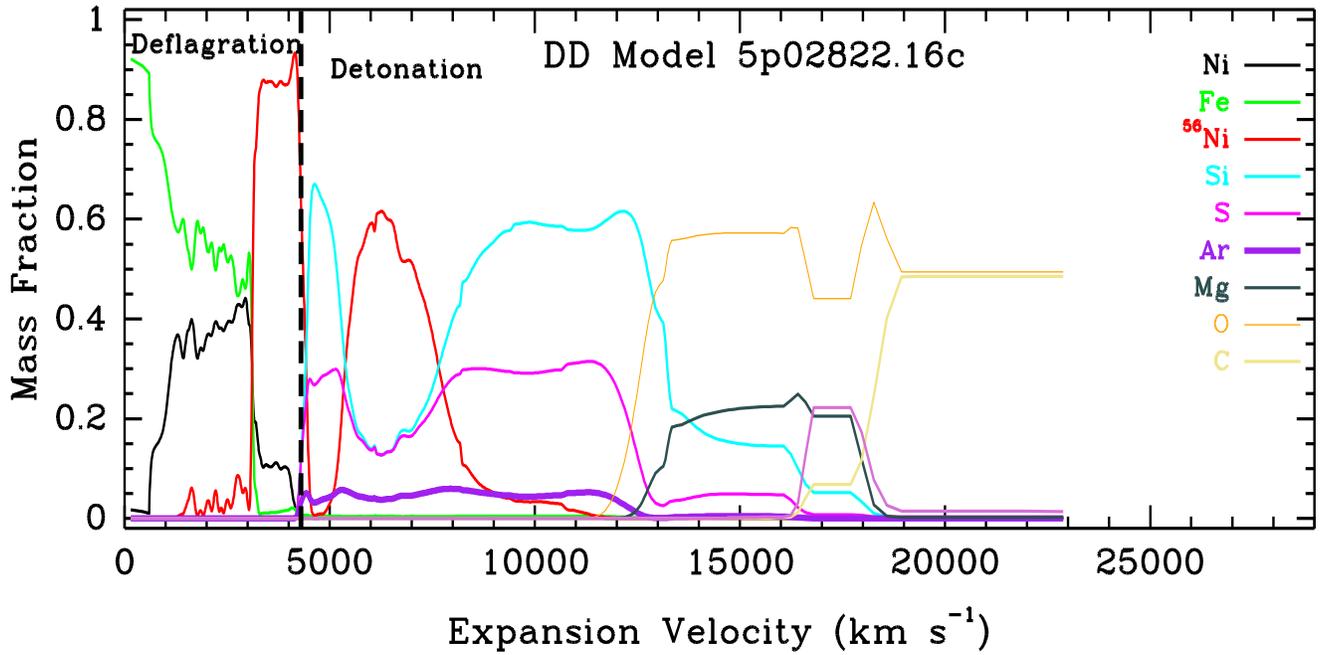}
\caption{
Abundances for the 5p02822.16 delayed detonation model (H02) plotted versus 
expansion velocity.  For simplicity, many of the less abundant species have not 
been plotted.  The vertical dashed line represents the approximate location of 
the transition from deflagration to detonation.  Material to the left of this
line is formed during the deflagration phase, while material to the right is 
the product of detonation.  Note that the small zone of Si/S-rich material near
5000 \kms\ is an artifact of the imposed spherical symmetry of this model and 
is not seen in multi-dimensional DD models.
\label{fig17}}
\end{figure*}

Indeed, the observed radial extent of these layers in SN~2005df matches quite
well to the predicted velocities from the 5p02822.16 DD model (see
Fig.~\ref{fig17}).  In this model, the Ar runs from $\sim 5000$ to 12000 \kms,
the radioactive \nicrad\ zone from  $\sim3000$ to 8000 \kms, and the stable Ni
in the inner $\sim 3000$~\kms, with a lower abundance ``tail'' out to $\sim
4000$~\kms(H02).

However, this innermost radial structure poses a bit of a problem even for DD
models.  The high-density electron capture burning in the H02 DD models all
takes place during the deflagration phase, and thus should be subject to the
same instabilities and mixing suggested by the 3-D deflagration models.  (The
fact that the H02 DD models do not show this is due to the fact that the
deflagrations are calculated in 1-D rather than 3-D.)  In the 5p02822.16 model
the detonation begins at a velocity of 4000--5000~\kms\ (marked as a vertical
dashed line in Fig.~\ref{fig17}), so the layers below this velocity should show
signs of mixing.  In SN~2005df, the Co does not appear to extend much below
$\sim 2500$~\kms, and the Ni is confined to a bubble of about $\sim 4000$~\kms\
in radius.  In fact, the only thing that suggests any sort of mixing in the
inner regions is the offset of the Ni from zero velocity, but this seems less
like the highly mixed structures of 3-D deflagration models
\citep{gamezo04,roepke06}, and (qualitatively at least) more like the buoyantly
rising bubble of the ``confined detonation'' scenario \citep{plewa04}.  

This is not the first set of observations that has failed to find evidence for
this mixing in the innermost ejecta of SNe~Ia, although it is probably the
clearest and most direct constraint.  Spherical DD models could reproduce the
NIR spectral evolution of the subluminous SN 1999by quite well.  However,
mixing the core of these models to simulate mixing during the deflagration
destroyed the good match to the observations (H02).   Support for a chemically
stratified ejecta structure was also found by \citet{badenes06}, in their
analysis of the X-ray spectra of the Tycho supernova remnant (SN 1572). Again, 
they found that the X-ray spectrum of Tycho could be quite well reproduced 
by modelling the interaction of the ejecta from a 1-D delayed detonation model
with the surrounding interstellar medium.

Late-time NIR observations of some SNe Ia show flat-topped boxy
[\ion{Fe}{2}] line profiles, indicative of a central hole in the distribution
of radioactive ejecta \citep{hoflich04,motohara06}.
Such a radiation hole can be explained as the result of nuclear burning at very
high densities during the deflagration, which produces stable Fe and Ni isotopes
via electron capture (EC) rather than radioactive \nicrad.  Again, however, if
the central region is mixed, then the non-radioactive electron-capture products
will be mixed with radioactive ejecta and a central radiation hole would not be
observed.

Possible alternative explanations for this observed central hole in the
radioactive ejecta remain unconvincing.  Some 3-D deflagration models
\citep[e.g.][]{roepke05} predict a large fraction of unburned material at low
velocities.  If the chemical differentiation is strong enough, this might
produce a central hole in the radioactive ejecta and thus the boxy line
profiles.  However, such a model would not explain the large mass of
low-velocity stable Ni seen in SN~2005df. Furthermore, models of the nebular
spectra based on the \citet{roepke05} explosion models predict strong carbon
and oxygen emission which is not seen in the late-time spectra of SNe Ia
\citep{kozma05}.  Interaction with the secondary star in a single-degenerate
scenario might produce a nearly conical hole in the ejecta
\citep{marietta00,kasen04}, but this will effect all layers of the ejecta
equally, or nearly so, and will not produce a central hole in only in the
innermost layers.  (However it is possible that this interaction might could
perhaps explain some of the bulk kinematic offsets seen in the late-time NIR
spectra of some SNe Ia as reported by \citealt{hoflich04, motohara06}.)

The MIR observations support the EC radiation hole explanation
for the boxy [\ion{Fe}{2}] line profiles.  In SN~2005df, we see direct evidence
that a significant amount of stable nickel is formed in the innermost ejecta,
which is confined to small radial velocities, and that there does appear to be
a hole in the center of the distribution of radioactive ejecta.  (These do not
quite coincide in velocity space since the Co distribution is centered, while
the Ni is slightly redshifted.)  Furthermore, the [\ion{Co}{3}] in SN~2003hv
is significantly shifted to the blue, and matches the observed blueshift of
the NIR [\ion{Fe}{2}] line \citep{motohara06}, which clearly supports
the notion that the late-time NIR [\ion{Fe}{2}] lines can be used to infer the
distribution of radioactive ejecta.

This observed lack of mixing has now been established in at least four SNe~Ia
(SN~1999by, SN~2003du, SN~2003hv, and SN~2005df), and as a result of three
entirely independent lines of inquiry (photospheric epoch NIR spectra,
late-time [\ion{Fe}{2}] line profiles, and mid-IR line profiles.)  
Furthermore,
this constraint does not simply come down on the side of detonations rather
than deflagrations. We are observing a lack of mixing in the innermost ejecta,
which burns as a deflagration in both the DD and pure deflagration scenarios.
This seems to suggest that something important is still missing from our
understanding of SN~Ia physics. Maybe something other than a deflagration is
responsible for the pre-expansion of the WD progenitor before a detonation.
Alternatively, perhaps there is some key piece of physics missing from the 3-D
deflagration models which maintains the radial structure of the 1-D models.

\section{Summary}
We have presented the first two mid-IR spectra of SNe~Ia.  
Although such observations are currently still quite difficult, requiring both
nearby SNe and space-based instruments, the mid-IR is certainly a rich
source of important data on these events.

The observed line profiles of SN 2005df show clear evidence of a highly
stratified chemical structure, with a small Ni-rich zone near the center, and
progressively larger surrounding \cobrad\ and Ar layers.  That the innermost
regions of the ejecta are so stratified seems to strongly support a detonation
scenario. Such a layered structure is incompatible with the highly mixed
chemical structures predicted by current 3-D deflagration models.  Furthermore,
this layered chemical structure persists even in the innermost ejecta, which
should contain the products of deflagration burning in both the DD and
pure-deflagration scenario, and suggests that a key piece of physics may still
be missing from the multi-dimensional models.

The line profiles of both SN~2003hv and SN~2005df also show signs of deviations
from spherical symmetry.  The [\ion{Co}{3}] line in SN 2003hv was blueshifted
by $\sim 3000$ \kms, consistent with the NIR [\ion{Fe}{2}] lines
\citep{motohara06}.  Such a kinematic offset in the radioactive ejecta can
perhaps be explained as the result of an off-center delayed-detonation. 

In contrast, the [\ion{Co}{3}] line in SN~2005df shows no significant kinematic 
offset, although there is an offset in the distribution of stable nickel in 
SN~2005df.  This cannot be explained with an off-center detonation, as the 
stable Ni is formed via electron capture during the deflagration phase.  It is
perhaps somewhat reminiscent of the buoyantly rising bubble in the ``confined
deflagration'' scenario.

The observed Ar line profiles in SN~2005df require an emission distribution 
that deviates significantly from spherical symmetry, especially for 
[\ion{Ar}{2}].  The implied asymmetry is much more pronounced than that seen in 
the calcium and iron distribution in SN~1885 \citep{fesen06}, and is much 
larger than the $\sim20\%$ asphericity implied by the strong continuum 
polarization in SN~1999by \citep{howell01}. The difference between the 
[\ion{Ar}{2}] and [\ion{Ar}{3}] line profiles suggests that much of this may be 
due to an asymmetric ionization structure rather than an asymmetric chemical 
structure.  

Our prolate Ar emission model for these lines is somewhat similar
to the predictions of off-center DD models, although currently these only 
produce a single Ar peak.  Currently these models also still assume a 
spherically symmetric deflagration, and therefore are also incompatible with
the observed Ni offset in SN~2005df.  It remains to be seen whether relaxing
this symmetry in an off-center DD model would produce a second Ar peak.  

On the
other hand, there is little sign of asymmetry in the radioactive \cobrad\ in 
SN~2005df, so it may be difficult to explain all three features in an off-center
DD scenario.  Indeed, we have also shown that the Ar lines can be reproduced in
a more ring-shaped geometry, although it requires a very flat distribution of 
[\ion{Ar}{2}] emission.

Finally, we have used the observed [\ion{Co}{3}] emission to infer \nicrad\
yields for both SNe using an LTE approximation, and assuming that the dominant
cobalt species is Co$^{2+}$.  For SN~2003hv this yields an estimate of
$M_{^{56}Ni}\approx 0.5 M_\sun$, although the derived electron density in the Co
zone may be well below the critical density for this SN. In contrast, for
SN~2005df, we find a low radioactive \nicrad\ mass, $M_{^{56}Ni} \approx 0.13$--$0.22
M_\sun$, and in this case the electron density should be near or above the
critical density for Co$^{2+}$ emission, so the LTE estimate should be
reasonably accurate.  Furthermore, it seems unlikely that other ionization
species could add enough to bring this up to the nominal
$M_{^{56}Ni}\sim0.5$--$0.6 M_\sun$ level.  SN~2005df appears to have been
subluminous, and so this low \nicrad\ yield seems to be direct evidence to
support the notion that the luminosity of SNe~Ia is primarily determined by the
\nicrad\ mass.

\acknowledgements
This work is based on observations made with the {\it Spitzer Space Telescope}, 
which is operated by the Jet Propulsion Laboratory, California Institute of 
Technology under NASA contract 1407. Financial support for the research
is provided by NASA/{\it Spitzer} grant GO-20256. C.L.G. is grateful for 
UK PPARC grant PPA/G/S/2003/00040.  J.C.W. is supported through NSF grant AST--0406740.  S.M. 
acknowledges financial support from the European Heads of Research Councils and
European Science Foundation EURYI Awards scheme.

\clearpage
\begin{deluxetable}{cccccccc}
\tablecaption{SN 2005df MIR Photometry\label{mir_phot}}
\tablewidth{0in}
\tablehead{
\colhead{JD Obs.} & \colhead{Epoch\tablenotemark{a}} & 
\multicolumn{6}{c}{$F_\nu$ (mJy)} \\
\cline{3-8}
\colhead{(245000+)} & \colhead{(days)} & 
\colhead{3.6~\micron\tablenotemark{b}} &
\colhead{4.5~\micron\tablenotemark{c}} &
\colhead{5.8~\micron\tablenotemark{d}} &
\colhead{8.0~\micron\tablenotemark{e}} &
\colhead{16~\micron\tablenotemark{f}} &
\colhead{22~\micron\tablenotemark{g}} 
}
\startdata
3676.5	& 93 & $0.36 \pm 0.03$ & $0.08 \pm 0.03$ & $0.11 \pm 0.07$ & 
$0.78 \pm 0.22$ & \nodata	& \nodata \\
3689.1 & 105 & \nodata & \nodata & \nodata & \nodata & $1.18 \pm 0.52$ 
& \nodata \\
3763.1 & 179 & \nodata & \nodata & \nodata & \nodata & $0.95 \pm 0.54$ 
& \nodata \\
3774.5 & 191 & $0.10 \pm 0.03$ & $0.03 \pm 0.02$ & $< 0.07$ & 
$0.68 \pm 0.17$ & \nodata	& \nodata \\
3952.6 & 369 & \nodata & \nodata & \nodata & \nodata & $< 0.31$ & $<0.92$\\ 
3955.6 & 372 & $< 0.03$ & $< 0.02$ & $< 0.11$ & $0.40 \pm 0.16$ &
\nodata & \nodata 
\enddata
\tablenotetext{a}{Assuming explosion occurred on 2 Aug. 2006}
\tablenotetext{b}{\textit{IRAC} Ch. 1}
\tablenotetext{c}{\textit{IRAC} Ch. 2}
\tablenotetext{d}{\textit{IRAC} Ch. 3}
\tablenotetext{d}{\textit{IRAC} Ch. 4}
\tablenotetext{f}{\textit{PUI} ``Blue''}
\tablenotetext{g}{\textit{PUI} ``Red''}
\end{deluxetable}

\begin{deluxetable}{cccccc}
\tablecaption{SN 2005df Emission Lines\label{lines}}
\tablewidth{0in}
\tablehead{
\colhead{$\lambda_{peak}$(\micron)} & \colhead{Species} & \colhead{Transition}& 
\colhead{$\lambda_{rest}$(\micron)} & \colhead{A$_{ki}$ (s$^{-1}$)} & 
\colhead{Line Flux\tablenotemark{a}}
}
\startdata
6.6 & [\ion{Ni}{2}] & $3d^9~^2D_{5/2}$--$3d^9~^2D_{3/2}$ & 6.636 & 
$5.5\times10^{-2}$ \tablenotemark{c} & $ <5$\tablenotemark{b} \\
6.91, 7.16 & [\ion{Ar}{2}] & $3p^5~^2P^o_{3/2}$--$3p^5~^2P^o_{1/2}$ & 6.985 & 
$5.28\times10^{-2}$ \tablenotemark{d} & $28\pm3$ \\
7.40 & [\ion{Ni}{3}] & $3d^8~^3F_4$--$3d^8~^3F_3$ & 7.349 & 
$6.5\times 10^{-2}$ \tablenotemark{d} &$6\pm2$\tablenotemark{b} \\
7.7--8.1 & SiO & $\Delta_v=1$ & $\sim 7.5$--8.0 & & $\sim 4$\\
8.49 & [\ion{Ni}{4}] & $3d^7~^4F_{9/2}$--$3d^7~^4F_{7/2}$ & 8.405 & 
$5.67\times 10^{-2}$ \tablenotemark{e} &$7\pm2$\tablenotemark{b} \\
8.90, 9.20 & [\ion{Ar}{3}] & $3p^4~^3P_2$--$3p^4~^3P_1$ & 8.991 & 
$3.1\times 10^{-2}$ \tablenotemark{f} &$23\pm2$ \\
10.60	& [\ion{S}{4}] & $3p~^2P^o_{1/2}$--$3p~^2P^o_{3/2}$ & 10.510 & 
$7.7\times 10^{-3}$ \tablenotemark{f} &$11\pm2$\\
 & [\ion{Co}{2}] & $3d^8~a^3F_4$--$3d^8~a^3F_3$ & 10.521 & 
 $2.21\times 10^{-2}$ \tablenotemark{g} \\
11.14 & [\ion{Ni}{3}] & $3d^8~^3F_3$--$3d^8~^3F_2$ & 11.002 &
$2.7\times 10^{-2}$ \tablenotemark{d} &$3.5\pm0.5$\\
11.99 & [\ion{Co}{3}] & $3d^7~a^4F_{9/2}$--$3d^7~a^4F_{7/2}$ & 11.888 & 
$2.01\times 10^{-2}$ \tablenotemark{d} &$31\pm2$ \\
12.5 & ? & & & &$\sim 2$ \\
12.8 & ? & & & &$\sim 2$\\
14.6 & ? & & & &$\sim 2$
\enddata
\tablenotetext{a}{In units of $10^{-15}$ erg cm$^{-2}$ s$^{-1}$.}
\tablenotetext{b}{Flux uncertain: line is blended with stronger feature.}
\tablenotetext{c}{\citet{quinet96}.}
\tablenotetext{d}{\citet{nussbaumer88}.}
\tablenotetext{e}{\citet{melendez05}.}
\tablenotetext{f}{NIST Atomic Spectra Database (V3.0.3)
\url{http://physics.nist.gov/PhysRefData/ASD/index.html .}}
\tablenotetext{g}{\citet{quinet98}.}
\end{deluxetable}

\clearpage
\begin{deluxetable}{ccccccccccc}
\tabletypesize{\footnotesize}
\tablecaption{Derived Abundances from Mid-IR Emission Lines\label{quantities}}
\tablewidth{0in}
\tablehead{
\multicolumn{2}{c}{Line ID} & \colhead{Obs. Mass (M$_\sun$)} & 
\multicolumn{8}{c}{LTE Inferred Mass\tablenotemark{a}(M$_\sun$) and $\tau_\nu$\tablenotemark{b}}\\
\cline{4-11}
& & & \multicolumn{2}{c}{1000 K} & \colhead{} & \multicolumn{2}{c}{6000 K} & \colhead{} & 
\multicolumn{2}{c}{10000 K} 
}
\startdata
\sidehead{SN~2005df\tablenotemark{c}}
~ [\ion{Ni}{2}] & \phn6.636 \micron & $<6.1\times 10^{-4}$ & $<8.7\times 10^{-3}$ & $<1.6$ & &
$<2.7\times 10^{-3}$ & $<0.1$ & &  $<4.1\times 10^{-3}$ & $<0.06$ \\
~ [\ion{Ni}{3}] & \phn7.439 \micron & \phm{$<$}$7.0 \times 10^{-4}$ & \phm{$<$} $7.1\times 10^{-3}$ & 
\phm{$<$}$2.4$ & & \phm{$<$}$2.4\times 10^{-3}$ & \phm{$<$}0.1 & & \phm{$<$}$2.7\times 10^{-3}$ &
\phm{$<$}0.08\\
~ [\ion{Ni}{3}] & 11.002 \micron & \phm{$<$}$1.5\times 10^{-3}$ & \phm{$<$}$7.8\times 10^{-2}$ & 
\phm{$<$}3.0 & & \phm{$<$}$8.8\times 10^{-3}$ & \phm{$<$}0.2 & & \phm{$<$}$9.0\times 10^{-3}$ & 
\phm{$<$}0.2\\
~ [\ion{Ni}{4}] & 8.406 \micron & \phm{$<$}$1.1\times 10^{-3}$ & 
\phm{$<$}$8.7\times 10^{-3}$ & \phm{$<$}3.6 & & \phm{$<$}$3.9\times 10^{-3}$ & 
\phm{$<$}0.2 & & \phm{$<$}$4.3\times 10^{-3}$ & 
\phm{$<$}0.1\\
\tableline
& & & \multicolumn{2}{c}{3000 K} & & \multicolumn{2}{c}{6000 K} & & \multicolumn{2}{c}{10000 K}\\
\tableline
~ [\ion{Co}{3}] & 11.888 \micron & \phm{$<$}$1.9\times 10^{-2}$ & 
\phm{$<$}$6.8\times 10^{-2}$ & \phm{$<$}0.8 & & \phm{$<$}$6.6\times 10^{-2}$ & 
\phm{$<$}0.4 & & \phm{$<$}$7.5\times 10^{-2}$ & 
\phm{$<$}0.2\\
~ [\ion{Ar}{2}] & 6.985 \micron & \phm{$<$}$2.4\times 10^{-3}$ & 
\phm{$<$}$1.2\times 10^{-2}$ & \phm{$<$}0.2 & & \phm{$<$}$9.0\times 10^{-3}$ & 
\phm{$<$}0.1 & & \phm{$<$}$8.2\times 10^{-3}$ & 
\phm{$<$}0.06\\
~ [\ion{Ar}{3}] & 8.991 \micron & \phm{$<$}$4.3\times 10^{-3}$ & 
\phm{$<$}$1.7\times 10^{-2}$ & \phm{$<$}0.2 & & \phm{$<$}$1.5\times 10^{-2}$ & 
\phm{$<$}0.1 & & \phm{$<$}$1.5\times 10^{-2}$ & 
\phm{$<$}0.06\\

\sidehead{SN~2003hv}
~ [\ion{Co}{3}] & 11.888 \micron & \phm{$<$}$4.5\times 10^{-3}$ & 
\phm{$<$}$1.7\times 10^{-2}$ & \phm{$<$}0.03 & & \phm{$<$}$1.7\times 10^{-2}$ & 
\phm{$<$}0.01 & & \phm{$<$}$1.8\times 10^{-2}$ & 
\phm{$<$}0.007
\enddata
\tablenotetext{a}{Using atomic partition functions of \citet{halenka01} for Ni;
\citet{halenka89} for Co; \citet{irwin81} for Ar.}
\tablenotetext{b}{Assuming $v_{max}$=4000 km s$^{-1}$ for Ni; 
$v_{min}$=2500 km~s$^{-1}$ and $v_{max}$=8000 km s$^{-1}$ for Co.
For Ar, the optical depths assume the density distribution of
the prolate line profile models.}
\tablenotetext{c}{SN~2005df abundances and optical depths calculated with an adopted distance of 
18.7 Mpc.}
\end{deluxetable}

\clearpage

\clearpage

\appendix
\section{Calculation of Nebular Emission-Line Profiles}
For nebular supernova emission with negligible sources of continuum opacity and
background emission, line profiles can be calculated in the Sobolev 
approximation by adding up the contributions from differential volume elements
in the SN envelope.  Adopting the notation of \citet{shu91}, the contribution to
the luminosity emitted toward the line of sight $\mathbf{\hat{k}}$ from 
material with specific intensity $I_\nu$ moving at (large) velocity 
$\mathbf{u}$ is given by
\begin{equation}
dL_\nu(\mathbf{\hat{k}})=4 \pi I_\nu \left(\frac{\nu_0}{c}\right) 
\mathbf{\hat{k}} \cdot \mathbf{\nabla(\hat{k}} \cdot \mathbf{u}) 
\left(\frac{dV}{d\nu}\right).
\end{equation}
For a homologously expanding envelope with $\mathbf{u}=\frac{R}{t}
\mathbf{\hat{r}}$, we have $\mathbf{\hat{k}} \cdot \mathbf{\nabla(\hat{k}} \cdot 
\mathbf{u})=t^{-1}$.  Assuming that the line emission is intrinsically narrow,
and placing the line of sight along the $x$ axis, then line emission with a 
rest-frame frequency $\nu_0$ will only contribute to the line profile at 
observed frequency $\nu$ if $\nu=\nu_0\left[1 + \frac{v_x}{c}\right]=
\nu_0\left[1 + \frac{x}{ct}\right]$.  Thus
\begin{equation}
dL_\nu(x)=4 \pi I_\nu \left(\frac{\nu_0}{c}\right) t^{-1} 
\left(\frac{dV}{dx}\right) \left(\frac{dx}{d\nu}\right)=
4 \pi I_\nu \left(\frac{dV}{dx}\right). \label{eqa2}
\end{equation}
In other words, the luminosity of the line profile at a given observed radial
velocity $v_x=x/t$ can be calculated by simply integrating the specific 
intensity of the emitting material in planar volume elements perpendicular to
the line of sight.  Thus, for a given spatial distribution of $I_\nu$, the line 
profile is, in principle, easy to calculate.  

In practice, however, supernovae are easier to describe as close to spherical 
or axisymmetric, so it is useful to convert this integration over a plane 
perpendicular to the line of sight, to an integration over spherical 
coordinates centered on the supernova and inclined at an angle $i$ to the line 
of sight (such that $i=0$ implies viewing the supernova ``pole-on'' and 
$i=90\degr$ corresponds to viewing the supernova equatorially or ``edge-on'').
In general, the specific intensity distribution can be decomposed into the 
product of three components,
\begin{equation}
I_\nu(r,\psi,\phi)=\rho(r,\psi,\phi) \Psi(\psi,\phi) \Phi(\phi), \label{eqa3}
\end{equation}
where $r$ and $\phi$ are the standard radial and longitudinal coordinates and 
$\psi$ is simply the complement of the standard latitudinal coordinate $\theta$.
For highly symmetrical distributions, the distribution will decouple further
with $\Psi(\psi,\phi)=\Psi(\psi)$ for axisymmetric distributions, and 
$\rho(r,\psi,\phi)=\rho(r)$ for spherical symmetry.  (Note: we chose to use 
$\psi$ rather than $\theta$ because it slightly simplifies the case of 
equatorial viewing.  This derivation can be relatively easily converted to the 
standard spherical coordinates by a judicious exchange of sines and cosines and 
swapping integration limits.)

It is useful to think of the differential volume elements as circular rings of 
constant $r$ and $\psi$.  Each ring will have a radius $R=r \cos \psi$, 
inclination $i$, and the $x$ coordinate of the center will be at $x_0=r \sin \psi
\cos i$.  For a given radial velocity $v_r=x/t$, the ring will only contribute 
the line profile at (at most) two points $\phi=\pm\phi'$ (except for the 
special case of $i=0$), where $\phi'=\cos^{-1}\left[\frac{(x-x_0)}{r \cos \psi
\sin i}\right]$.  The differential volume element $dV=r^2 \cos \psi dr d\psi 
d\phi$, and so combining Equations \ref{eqa2} and \ref{eqa3}, we have
\begin{eqnarray*}
dL_\nu(x)&=&4 \pi \rho(r,\psi,\phi) \Psi(\psi,\phi) \Phi(\phi) 
[\delta(\phi-\phi') + \delta(\phi + \phi')]
r^2 \cos \psi dr d\psi \left(\frac{d\phi}{dx}\right)  \\
&=& 4 \pi \rho(r,\psi,\phi) \Psi(\psi,\phi) \Phi(\phi) 
[\delta(\phi-\phi') + \delta(\phi + \phi')]
\frac{r^2 \cos \psi dr d\psi}{\sqrt{r^2 \cos^2 \psi \sin^2 i - (x-x_0)^2}}. 
\end{eqnarray*}
The total contribution to the line flux at a given radial velocity can thus be 
found by integrating over those rings that intersect the $x=v/t$ plane. Hence,
\begin{equation}
L_\nu(x=v_r/t)=\int_{x}^{v_{max}/t} r^2 \rho(r) 
\int_{\psi_{min}(r,x)}^{\psi_{max}(r,x)} \Psi(r,\psi) \Phi^*(x,\psi,r)
\frac{\cos \psi d\psi dr}{\sqrt{r^2 \cos^2 \psi \sin^2 i - (x-x_0(r,\psi))^2}},
\label{eqa4}
\end{equation}
where 
\begin{displaymath}
\Phi^*(x,\psi,r)=\left[\Phi(\phi'(x,\psi,r)) + \Phi(-\phi'(x,\psi,r))\right]
\end{displaymath} for $i\neq0$ and 
\begin{displaymath}
\Phi^*(x,\psi,r)=\int_0^{2 \pi} \Phi(\phi)d\phi
\end{displaymath} for $i=0$.
The latitudinal limits of integration are more complicated, and for a given
$r$ depend on the value of $x$ relative to $x_N=r \cos i$ and $x_S=-r \cos i$,
the $x$-coordinates of the $\psi=\pm90\degr$ poles on a sphere of radius $r$.
For a given value of $r$, the latitudinal integration limits are
\begin{eqnarray*}
\mbox{For } x_N \leq x \leq r \mbox{:} & \\
& \psi_{min}(r,x)=\frac{\pi}{2} - \alpha - i\\
& \psi_{max}(r,x)=\frac{\pi}{2} + \alpha - i\\
\mbox{For } 0 \leq x < x_N \mbox{:} & \\
& \psi_{min}(r,x)=\frac{\pi}{2} - \alpha - i\\
& \psi_{max}(r,x)=\frac{\pi}{2} - \alpha + i\\
\mbox{For } x_S \leq x < 0 \mbox{:} & \\
& \psi_{min}(r,x)=-\frac{\pi}{2} + \alpha - i\\
& \psi_{max}(r,x)=-\frac{\pi}{2} + \alpha + i\\
\mbox{For } -r \leq x < x_S \mbox{:} & \\
& \psi_{min}(r,x)=-\frac{\pi}{2} - \alpha + i\\
& \psi_{max}(r,x)=-\frac{\pi}{2} + \alpha + i\\
\end{eqnarray*}
where $\alpha=\cos^{-1}(\frac{|x|}{r})$.

For the line profiles presented in this work, the radial, latitudinal, and 
longitudinal specific intensity distribution functions (Eqn~\ref{eqa3}) were 
coded in Python and Equation~\ref{eqa4}\ numerically integrated using the 
{\tt quadpack} algorithms in the 
SciPy\footnote{\url{http://www.scipy.org/SciPy}} software package.  
\begin{figure*}[t]
\includegraphics[scale=0.65,angle=270,clip=True,trim=165bp 0bp 165bp 0bp]{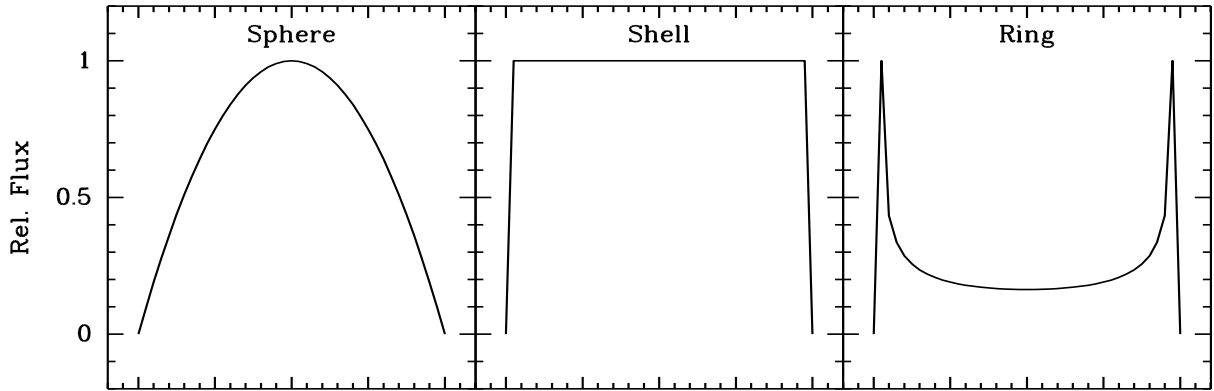}
\caption{Calculated emission-line profiles for three simple emission geometries:
a uniform sphere (left), a uniform thin spherical shell (center), and a uniform
thin circular ring (right).
\label{example_profs}}
\end{figure*}
The resulting line profiles from simple emission geometries are shown in 
Figure~\ref{example_profs}.  A homogeneous sphere of emission produces the 
well-known parabolic line-profile shape, while a thin, uniformly emitting spherical
shell produces a ``top-hat'' line profile, and a thin, uniformly emitting
ring produces a forked profile [$F\propto (1-(v/v_{max})^2)^{-1/2}$].  Since
the volume integration in Equation~\ref{eqa4} is ``additive,'' it is often 
useful to think of the emission from more complicated geometries as linear
combinations of simpler emission geometries.  Thus a thick sphere of emission 
with a central hole can be built up either by adding ``top-hat'' line profiles 
from progressively larger shells, or by subtracting the line profile from a 
small homogeneous sphere from that of a larger homogeneous sphere.  In both cases
one ends up with a line profile that is flat (shell-like) in the core, and 
parabolic (sphere-like) in the wings.
\end{document}